\documentclass[preprint,5p]{elsarticle}
\usepackage{graphicx}
\usepackage{caption}
\usepackage{subcaption}
\usepackage{amsmath}
\usepackage{siunitx}
\usepackage{booktabs}
\usepackage{multirow}
\usepackage[hyphens]{url}
\usepackage[colorlinks]{hyperref}
\usepackage[noabbrev,capitalise,nameinlink]{cleveref}
\usepackage{array}
\usepackage{bm}
\usepackage{fancyhdr}

\biboptions{square,comma,numbers,sort&compress}

\DeclareSIUnit{\nothing}{\relax} 

\begin{document}

\journal{Acta Astronautica}

\begin{frontmatter}

\title{System Modelling of Very Low Earth Orbit Satellites for Earth Observation}

\author[UoM]{N.H.~Crisp}\corref{cor1}
\ead{nicholas.crisp@manchester.ac.uk}

\author[UoM]{P.C.E.~Roberts}
\author[Stuttgart]{F.~Romano}
\author[UoM]{K.L.~Smith}
\author[UoM]{V.T.A.~Oiko}
\author[Deimos]{V.~Sulliotti-Linner}
\author[Gomspace]{V.~Hanessian}
\author[Stuttgart]{G.H.~Herdrich}
\author[UPC]{D.~Garc\'{i}a-Almi\~{n}ana}
\author[MSSL]{D.~Kataria}
\author[Euroconsult]{S.~Seminari}

\cortext[cor1]{Corresponding author.}

\address[UoM]{The University of Manchester, Oxford Road, Manchester, M13~9PL, United Kingdom}
\address[Deimos]{Elecnor Deimos Satellite Systems, Calle Francia 9, 13500 Puertollano, Spain}
\address[Gomspace]{GomSpace A/S, Langagervej~6, 9220 Aalborg East, Denmark}
\address[Stuttgart]{Institute of Space Systems (IRS), University of Stuttgart, Pfaffenwaldring~29, 70569 Stuttgart, Germany}
\address[UPC]{UPC-BarcelonaTECH, Carrer de Colom~11, 08222 Terrassa, Barcelona, Spain}
\address[MSSL]{Mullard Space Science Laboratory, University College London, Holmbury St. Mary, Dorking, RH5~6NT, United Kingdom}
\address[Euroconsult]{Euroconsult, 86~Boulevard de Sébastopol, 75003 Paris, France}

\begin{abstract}
The operation of satellites in very low Earth orbit (VLEO) has been linked to a variety of benefits to both the spacecraft platform and mission design. Critically, for Earth observation (EO) missions a reduction in altitude can enable smaller and less powerful payloads to achieve the same performance as larger instruments or sensors at higher altitude, with significant benefits to the spacecraft design. As a result, renewed interest in the exploitation of these orbits has spurred the development of new technologies that have the potential to enable sustainable operations in this lower altitude range. In this paper, system models are developed for (i) novel materials that improve aerodynamic performance enabling reduced drag or increased lift production and resistance to atomic oxygen erosion and (ii) atmosphere-breathing electric propulsion (ABEP) for sustained drag compensation or mitigation in VLEO. Attitude and orbit control methods that can take advantage of the aerodynamic forces and torques in VLEO are also discussed. These system models are integrated into a framework for concept-level satellite design and this approach is used to explore the system-level trade-offs for future EO spacecraft enabled by these new technologies. A case-study presented for an optical very-high resolution spacecraft demonstrates the significant potential of reducing orbital altitude using these technologies and indicates possible savings of up to \SI{75}{\%} in system mass and over \SI{50}{\%} in development and manufacturing costs in comparison to current state-of-the-art missions. For a synthetic aperture radar (SAR) satellite, the reduction in mass and cost with altitude were shown to be smaller, though it was noted that currently available cost models do not capture recent commercial advancements in this segment. These results account for the additional propulsive and power requirements needed to sustain operations in VLEO and indicate that future EO missions could benefit significantly by operating in this altitude range. Furthermore, it is shown that only modest advancements in technologies already under development may begin to enable exploitation of this lower altitude range. In addition to the upstream benefits of reduced capital expense and a faster return on investment, lower costs and increased access to high quality observational data may also be passed to the downstream EO industry, with impact across a wide range of commercial, societal, and environmental application areas.
\end{abstract}

\begin{keyword}
Very low Earth orbit (VLEO); Atmosphere-breathing electric propulsion (ABEP); Aerodynamic control; Very-high resolution; Synthetic aperture radar (SAR); Earth observation (EO).
\end{keyword}

\end{frontmatter}

\section{Introduction} \label{S:Introduction}
Satellite-based Earth observation (EO) and remote sensing has evolved considerably in the six decades since its introduction and has become increasingly important as a field with global scientific significance, economic impact, and commercial value. In addition to improving our understanding of the Earth and its natural systems, key applications with societal and environmental impact include climate change and ecological monitoring, agriculture and food security, and disaster management and response \cite{Kansakar2016}. The commercial market for satellite-based EO data products has also grown considerably, principally serving sectors such as defence and security, infrastructure monitoring, land management and precision agriculture, and management of energy and mining activities. Revenue from these products is projected to increase from \$1.5~billion to \$2.4~billion in the decade to 2028 with increasing demand for higher resolution imagery \cite{Euroconsult2019}. Constellations of smaller satellites for EO have recently enabled and global coverage and more frequent revisit, contributing to increased availability and a reduction in price of medium and high resolution satellite imagery and resulting in significant growth in  downstream value-added services and analytics market segments. Further economic upside has been projected based on the development and growth of new application areas, for example in change-detection analytics serving the financial sector and the development of new location-based services \cite{Euroconsult2019}.

Operating satellites at lower orbital altitude has been proposed as a means to support the growth in this market by reducing both the cost of satellite development and launch and through the generation of higher resolution imagery \cite{Wertz2012a,Roberts2017,Roberts2019,Roberts2020}. Reducing orbital altitude below those conventionally used offers a number of individual benefits that together can significantly improve the design and performance of satellites, particularly for remote sensing and EO applications. These benefits can be summarised as follows \cite{Crisp2020}:
\begin{itemize}
\item Optical payloads increase in resolution or can reduce in aperture size resulting in either improved performance or smaller size and mass.
\item Radiometric performance also improves with reducing distance to the target or ground, allowing improvement of signal-to-noise ratio or reduced instrument sensitivity. This also applies to radar and communications payloads leading to improved link budgets, reduction in power, and smaller antenna areas.
\item Lower altitude orbits are naturally resilient to a build-up in debris due to the effects of drag and therefore have a lower risk of on-orbit collision.
\item The same effect of drag ensures that spacecraft are naturally disposed of quickly after their mission is complete or if they suffer a catastrophic failure. Additional deorbit devices are therefore not required, reducing system mass and complexity.
\item Launch vehicles can deliver a larger mass into lower altitude orbits, reducing the specific (per unit mass) launch cost and possibly providing greater versatility in launch options.
\item Mapping errors as a result of attitude determination and pointing accuracy are reduced at lower altitude, improving the geospatial accuracy of ground imagery and location-based services.
\item Use of commercial off-the-shelf (COTS) electronics components may be enabled as a result of reduced radiation exposure and dosing in the denser atmosphere, reducing cost and the need for redundancy.
\end{itemize}

However, despite these benefits, altitudes below \SI{450}{\kilo\meter}, known as very low Earth orbit (VLEO), have not been frequently utilised for commercial EO operations. Some military satellites, for example in the Keyhole program, have been reported to use eccentric orbits with very-low perigees to obtain high resolution surveillance imagery \cite{Richelson1984}. In recent times, VLEO has also been populated by the International Space Station (ISS), scientific satellites (e.g. GOCE and Grace), and small educational and technology development satellites (e.g. CubeSats). Early satellites of the Planet Labs Flock and Spire Lemur constellations were also deployed into the VLEO altitude range from the ISS. However, more recent operations have generally seen replacements launched into higher altitude orbits. The reason for this sparse usage of VLEO, particularly for commercial operations, is largely due to the challenges of the environment that can significantly reduce the mission performance. Principally, the increased atmospheric density increases aerodynamic drag and causes a significant reduction in orbital lifetime without the use of propulsion. The presence of atomic oxygen further reduces aerodynamic performance and can erode and damage the external surfaces of spacecraft.

The recent development of a number of new technologies aims to address these challenges to enable the sustainable operation of spacecraft in VLEO. These technologies principally include: i) novel materials that reduce aerodynamic drag and are resistant to atomic oxygen erosion, ii) atmosphere-breathing electric propulsion (ABEP) for drag compensation or mitigation in VLEO, and iii) aerodynamic attitude and orbit control methods.

This paper aims to explore the use of these novel technologies being developed to enable operations at reduced orbital altitude and their impact on the design of future VLEO spacecraft for EO missions. To perform this investigation, system models incorporating VLEO technologies have been developed and integrated into a design framework for concept-level spacecraft design. Using the developed scheme, the system-level trade-offs associated with these technologies can be illustrated and their variation with orbital altitude understood. The effect of assumed performance on the system-level design can also be demonstrated, leading to insight into the requirements of these technologies for improving spacecraft design for VLEO. The results of these investigations will be used to inform the ongoing research in this area and the development of these technologies and future spacecraft operating in VLEO.

\subsection{The Very Low Earth Orbit (VLEO) Environment}
\begin{figure*}
	\centering
	\includegraphics[height=180mm,angle=270]{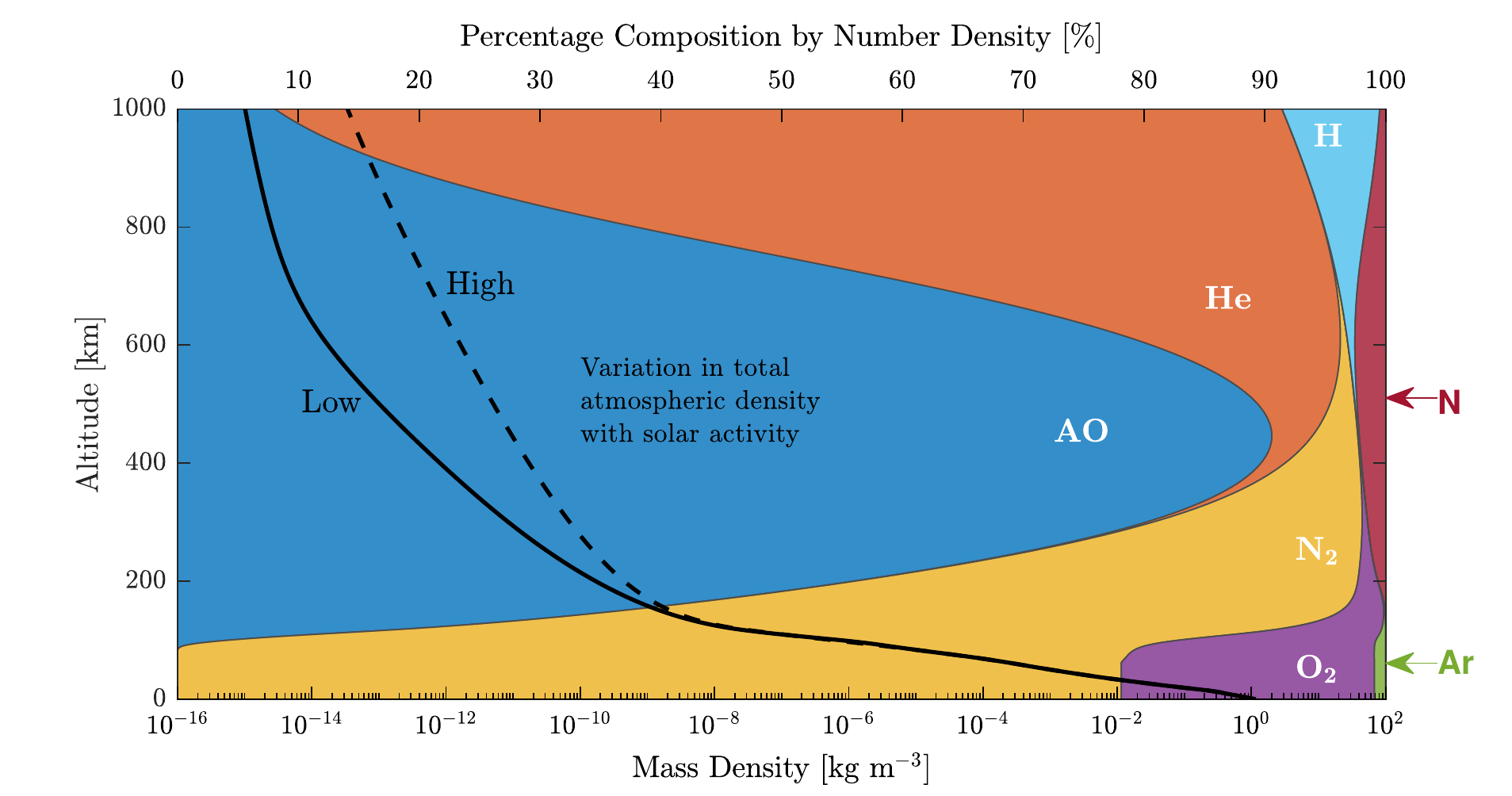}
	\caption{Representative variation of atmospheric density and composition with solar activity level and altitude in LEO, calculated using the NRLMSISE-00 model \cite{Picone2002}.}
	\label{F:VLEO_Environment}
	\end{figure*}

As altitude in low Earth orbit (LEO) is reduced the density of the residual atmosphere increases due to the effect of gravity and hydrostatic pressure as shown in \cref{F:VLEO_Environment}. The aerodynamic drag force experienced by objects in lower altitude orbits therefore increases, causing faster orbital decay and eventually atmospheric re-entry and demise. Without the use of a propulsion system that can compensate for or mitigate the experienced drag, this causes spacecraft at very low altitude to have significantly shorter orbital lifetimes compared to higher altitude orbits. However, in order to perform drag compensation or mitigation the spacecraft must be launched with or resupplied with sufficient propellant. Notable examples include the ISS, which maintains an approximately \SI{400}{\kilo\meter} orbit through propellant provided by frequent resupply missions, and GOCE that was able to sustain its orbit for over \SI{4}{years} at altitudes below \SI{280}{\kilo\meter} and as low as \SI{229}{\kilo\meter} using an highly capable ion propulsion system \cite{Steiger2014}.

Despite this increase in atmospheric density, the residual atmosphere in VLEO is still highly rarefied. Above altitudes of approximately \SI{150}{\kilo\meter} the flow condition for orbiting spacecraft of typical length scale ($\sim$\SI{1}{\meter}) is considered to be free molecular ($Kn \gg 10$) \cite{Sentman1961}. In the free molecular flow (FMF) regime aerodynamic forces are principally generated by the direct interaction between the particles in the flow and the external spacecraft surfaces as interactions between the constituent particles are infrequent by comparison and can therefore be neglected. These gas-surface interactions (GSIs) are dependent on the energy and momentum exchange that occurs between the incoming particle and the surface, and are known to vary depending on the properties of the incoming particles (mass, velocity, angle, temperature) surface characteristics (roughness, contamination, composition, and morphology) \cite{Moe1998}. The estimation of aerodynamic forces under these conditions can be calculated using a range of different GSI models, reviews of which are provided by \citet{MostazaPrieto2014} and \citet{Livadiotti2020}.

In combination with the increasing density and orbital velocity, the prevalence of highly-reactive atomic oxygen at these altitudes, shown in \cref{F:VLEO_Environment}, can also damage and erode materials on the external surfaces of spacecraft, degrading both the aerodynamic performance and affecting optical sensors, thermal coatings, and solar-cell cover-glass \cite{Banks2004}. In addition to this erosive damage, atomic oxygen also adsorbs to the external spacecraft surfaces causing surface contamination and further affecting the GSI parameters. For typical materials in the VLEO environment, diffuse reflection with complete thermal accommodation is generally assumed, primarily as a result of atomic oxygen erosion and surface adsorption \cite{Pardini2010}. As a result of this energy and momentum exchange, the main aerodynamic force generated is drag and is principally dependent the projected cross-sectional area rather than the more detailed spacecraft geometry. However, at higher altitude reduced atomic oxygen adsorption has been observed and incomplete energy accommodation and quasi-specular particle re-emission can be considered \cite{Moe1998,Macario-Rojas2018}. 

The atmospheric density experienced at a given altitude can also vary considerably, principally with geographic position (e.g. latitude and longitude), the day-night cycle, and the approximately 11-year cycle of solar intensity \cite{Emmert2015}. This variation from high to low solar activity level is shown in \cref{F:VLEO_Environment} and can be seen to span several orders of magnitude above approximately \SI{100}{\kilo\meter}. Similarly, the direction of the oncoming flow can vary as a result of atmospheric co-rotation and thermospheric winds. When combined with the external spacecraft geometry and configuration, these effects can generate disturbing forces and torques that can effect the satellite stability and pointing capability. For EO satellites with strict pointing and stability requirements, highly capable attitude actuators may be needed to compensate for these effects and ensure the ability to accurately point towards, track, and capture a given target.

\subsection{Emerging Technologies for VLEO}
A range of fundamental research and technology development is underway to address these critical challenges associated with the environment and enable sustained operations in VLEO.

Research into novel materials for orbital aerodynamics is aiming to identify those that are resistant to the detrimental effects of atomic oxygen and demonstrate specular or quasi-specular GSI properties. When combined with the design of the spacecraft external geometry (i.e. forward-facing surfaces oriented at shallow angles to the oncoming flow), materials with these properties would experience lower drag in orbit and would therefore increase the orbit lifetime. Alternatively, if oriented towards the normal to the oncoming flow, these materials increase the experienced drag and can be applied to the design of improved deorbit devices. These materials would also be able to generate lift forces of increased magnitude enabling novel methods of aerodynamic attitude and orbit control. 

The ongoing development of space propulsion systems is highly relevant to VLEO spacecraft. Efficiency increases of electric propulsion (EP) systems will reduce the power and propellant requirements for drag compensation or mitigation. Beyond this, the development of novel atmosphere-breathing electric propulsion (ABEP) systems would enable theoretically sustainable operation in VLEO, limited only by component lifetime. Such concepts propose the use of a forward-facing intake that collects propellant from the residual atmosphere and can subsequently be used by an electric thruster, eliminating the need for the spacecraft to carry or be launched with any propellant. The development of ABEP systems is currently focused on the design of intakes that can efficiently collect the required propellant from the atmosphere under the FMF conditions in VLEO \cite{Singh2015} and the development of electric thrusters that can flexibly utilise the range of different atmospheric gas constituents at varying density that will be collected, whilst remaining resilient to erosion or degradation by aggressive atmospheric species \cite{Schonherr2015}. Electrodeless and gridless thruster concepts that do not require direct contact with the atmospheric gas therefore demonstrate good potential for use in VLEO, for example the RF helicon-based inductive plasma thruster \cite{Romano2020}. Such a design benefits from the quasi-neutral plasma flow that eliminates the need for a neutralizer. For non-quasi-neutral thrusters, operation with the residual constituents of the atmosphere has not yet been solved.

The development of ABEP may also facilitate the use of novel orbit control methods that may help to overcome the drawbacks of VLEO with regards to increased revisit time and generally poor coverage properties. Methods for low-thrust trajectory design have been developed to reduce the overflight or revisit time of a set of given ground targets from low altitude orbits \citet{Guelman1999,Co2014,Guelman2019}. With ABEP, the constraint of limited propellant can be removed, and such control may be extended to include small out-of-plane manoeuvres and applied throughout the entirety of a mission. Progress towards methods for autonomous and real-time control may further facilitate their use in the operation of future VLEO constellations.

New methods of attitude and orbit control that utilise aerodynamic forces and torques are also being developed. Proposals for orbit control include constellation deployment \cite{Leppinen2016}, formation keeping \cite{Leonard1989,Traub2019a}, rendezvous \cite{Bevilacqua2008,DellElce2015}, inclination correction for sun-synchronous orbits \cite{VirgiliLlop2015}, and atmospheric re-entry interface targeting \cite{VirgiliLlop2015a,Omar2019}. Aerodynamics attitude control has also been proposed for pointing and momentum management manoeuvres that assist and reduce the requirements on traditional attitude actuators \cite{Gargasz2007,Auret2011,Mostaza-Prieto2017,Livadiotti2021}, with the potential for reducing system mass. Aerodynamic trim manoeuvres can also be considered to directly reject external disturbances, for example as a result of variations in the oncoming flow direction, solar radiation pressure, and residual magnetic dipole interactions. Such control may be particularly relevant at very low altitude where aerodynamic torques may cause the rapid saturation of momentum-based actuators. However, at present only orbit control methods utilising differential drag for collision avoidance \cite{Mishne2017,Huang2017} and constellation deployment/maintenance \cite{Maclay2005,Gangestad2013,Foster2018} have been demonstrated in-orbit. Similarly, aerodynamic attitude control in the form of trim and momentum management has only been demonstrated on MagSat \cite{Tossman1980,Stengle1980}). This is principally due to the low lift-to-drag ratio that is achievable with current materials and the high uncertainties associated with knowledge of the oncoming flow density and velocity, leading to low expectations on control performance. Without further generalised study, implementation, and demonstration of such control methods, modelling of their system-level requirements, trade-offs, and effects with respect to conceptual designs is difficult and is beyond the scope of this current study. As development continues and initial on-orbit demonstrations are performed, concepts and methods of aerodynamic control can be integrated into future studies of VLEO platform design. Key trade-offs associated with the implementation of aerodynamic control are expected to include actuator size, mass, and configuration, pointing performance and momentum control/dumping capability, and power requirements. In the wider system design, the corresponding reduction in traditional attitude or orbit control actuator requirements and the drag increment associated with the aerodynamic control implementation should also be considered.

\section{System Modelling for Reducing Altitude in LEO}
System modelling for satellites and orbital space systems has been developed over a number of years, the basics of which are well established in texts such as \citet{Wertz1999}, \citet{Wertz2011}, and \citet{Fortescue2011}. These system models allow engineers to investigate different designs and synthesise solutions in the early design process. Exploration of the design- and trade-spaces, often by parametric or optimisation processes, can also be performed to improve knowledge of the trade-offs between different input variables, subsystem specifications, and output parameters. This exploration can also facilitate the identification and selection of promising or optimal solutions that can subsequently be taken forward for further investigation and into more detailed design phases.

The selection, integration, and configuration of the system/subsystem models is specific to the mission, concept of operations, and application area of the system of interest. However, general approaches to complex and multidisciplinary systems engineering can be used to inform the development of an appropriate framework and the organisation of the contributing models \cite{Lambe2012}. Within the discipline of space systems, application of such system modelling is widespread with differing levels of integration and scope, e.g. at system level \cite{Mosher1999,Fukunaga1997,Barnhart2009a,Lowe2014,Hwang2014}, constellation design \cite{LeMoigne2017,Ferringer2006} and deployment \cite{DeWeck2004,Crisp2019}, propulsion systems \cite{Spangelo2014,Dono2018,Krejci2018}, and communications architecture \cite{DeWeck2002}. However, given the challenges of operating spacecraft in low altitude orbits, to date only a limited number of studies have approached the system modelling of spacecraft intended for operation in VLEO.

The design of remote-sensing satellites operating in lower altitude orbits was studied by \citet{Fearn2005} with a particular focus on drag compensation using EP. The system-level trades in propulsive, power system, and payload design with altitude were principally considered for a mission with a \SI{5}{year} lifetime. The design of a concept spacecraft providing \SI{1}{\meter} resolution imaging was presented with an operational altitude between \SIrange{250}{280}{\kilo\meter} and wet mass of less than \SI{300}{\kilo\gram}. Whilst the presented design suggested the near-term feasibility of such a concept, challenges such as the required solar array performance and sensitivity to on-orbit solar illumination were noted. 

A review of the benefits of moderately-elliptical very-low orbits (MEVLOs) was provided by \citet{Wertz2012a}. In this work the propellant required to maintain different low-perigee orbits was explored for the purpose of enabling high-resolution imaging from a smaller and cheaper satellite platform. A basic concept spacecraft called NanoEye was proposed with a wedge-shaped front geometry to reduce drag and a propulsion system to provide orbit maintenance and tasking capabilities. 

A later study by \citet{Shao2016} focused on more classical circular orbits and applied a performance-based cost modelling approach to investigate the trade-off between total system cost for an EO constellation and the operational orbit altitude. This study indicated that significant cost savings could be made for equivalent system performance (resolution and coverage) by reducing the orbital altitude below \SI{500}{\kilo\meter}. However, the modelling approach considered only conventional spacecraft designs and did not explore the impact of novel technologies that are currently being developed and are applicable to satellites operating in VLEO.

Concept missions for VLEO have also developed by design teams at Cranfield University \cite{VirgiliLlop2014a}. A satellite based on the SSTL-300 platform operating between \SIrange{300}{500}{\kilo\meter} was designed to be implemented in the near-term, providing enhanced resolution predominantly by reducing the orbital altitude of operation. Different propulsion systems were chosen depending on the selected altitude to provide a mission lifetime of \SI{5}{years}. A more ambitious spacecraft called THOR (Thermospheric Orbital Reconnaissance) was also developed to operate in a \SI{227}{\kilo\meter} sun-synchronous orbit (SSO) and provide \SI{0.15}{\meter} resolution optical imagery. Drag-compensation was provided by an EP system and the geometry was designed to reduce drag and provide natural aerostability. Internal steerable optics were considered to provide some off-nadir pointing performance. 

The ``Skimsat'' has been proposed by \emph{Thales Alenia Space} \cite{Bacon2017}. This concept is designed to be deployed into EO constellations in VLEO as low as \SI{160}{\kilo\meter} in altitude, benefiting from the reduction in payload size and mass with altitude. The spacecraft geometry suggests that it has been designed for operation in the aerodynamic environment of VLEO and remarks are made regarding use of ABEP for drag compensation. However, whilst some discussion of the key system trade-offs is provided, details of the underlying system design has not yet been presented. Similar concepts are also being developed by emerging companies such as \emph{EarthObservant} and \emph{Skeyeon} demonstrating a growing interest from industry in VLEO operations.

\citet{McCreary2019} presents a traditional systems engineering approach to the development of a satellite mission concept for operation in the high drag environment of VLEO. The initial needs analysis is performed principally for US scientific, military, and academic stakeholders and used to develop a sample set of system requirements. The use of both a fight-proven EP system with stored propellant and an ABEP system are considered to provide redundancy and allow for initial operational learning at the beginning of the mission before the ABEP system is given full responsibility for drag-compensation. A maximum cross-sectional area of \SI{1.5}{\meter\squared} and characteristically high drag coefficient of \num{2.7} are selected, but only solar minimum conditions are assumed for the estimation of atmospheric density. The solar arrays also do not appear to be included in the calculation of the aerodynamic drag. A fixed power efficiency of \SI{40}{\milli\newton\per\kilo\watt} is assumed for both the EP and ABEP systems and is not connected to the operational altitude, atmospheric intake characteristics, or thruster performance. System models from \emph{Space Mission Analysis and Design} \cite{Wertz1999} are used to perform the initial spacecraft sizing (wet mass of \SI{1138}{\kilo\gram} and solar array area of \SI{41.1}{\meter\squared}) and to generate an initial 3D model for the proposed spacecraft. 

The use of EP to enable new EO systems in at lower altitude has also recently been considered by \citet{Bertolucci2020}. In this work, the performance parameters and operational considerations for different conventional EP systems are analysed for reducing orbital altitude. Basic system models are also used estimate the remaining spacecraft size and mass at different altitude. A nominal drag coefficient of \num{2.2} is adopted and the associated cross-sectional area is determined only from the total mass and an assumed spacecraft density. The authors conclude that present EP systems could be used to perform drag compensation in the upper range of VLEO, resulting in satellite masses of between two and six times less for the same resolution as equivalent systems operating at a higher altitude. At lower altitude, it is noted that the propulsion system performance (i.e. specific impulse) becomes much more critical to successful operations.  

\subsection{Atmosphere-Breathing Electric Propulsion Concepts}
Work towards the development of ABEP systems has also often been accompanied by analysis of the performance of a combined intake and thruster with altitude. Concepts for the spacecraft that they may be initially demonstrated on are sometimes also presented. However, in most cases these designs are aimed only at technology demonstration and therefore do not feature payloads that would be required by commercially-focused satellites.

The Air Breathing Ion Engine (ABIE) concept \cite{Nishiyama2003,Hisamoto2012} is presented as an additional propulsion module that can be attached to the tail end of a long slender spacecraft where the atmospheric intake is mounted in an external concentric ring. The system design considers the variation of inlet-to-thruster area ratio and power that is needed to provide drag-compensation at different altitude. However, whilst solar electric power is assumed, the studies do not address the generation of this power (e.g. body-mounted or external arrays). A static drag coefficient of \num{2.0} is assumed and the cross-sectional area of the intake is neglected in the calculation of aerodynamic drag.

The ESA RAM-EP study \cite{DiCara2007} provides the first investigation of different thrusting strategies for drag-compensation (continuous or periodic) and analyses the system performance at varying altitude and under different solar activity conditions. The study also proposes a mission to demonstrate the performance of the propulsion system and to perform EO using a lidar payload. A mass of \SI{1000}{\kilo\gram} is targeted and a total cross-sectional area of \SI{1}{\meter\squared} is assumed with an intake-area of \SI{0.6}{\meter\squared}. A fixed drag coefficient of \num{2.0} is stated and the total required area of solar arrays calculated as \SI{19.7}{\meter\squared}. However, it is not established if the deployed solar arrays are considered in the calculation of the aerodynamic drag and therefore the required power.

System analysis for ABEP concepts is also presented by \citet{Schonherr2015} and \citet{Romano2018c} in order to inform the development of the RF Helicon-based IPT \cite{Romano2020}. A cross-sectional area of \SI{1}{\meter\squared} and a fixed drag coefficient of \num{2.2} were selected and it was assumed that the drag contribution of solar arrays could be neglected as they would be fixed parallel to the oncoming flow. It is also noted in this study that for the atmospheric intakes considered, a design loop is present between the intake area, thruster inlet area, mass-flow rate, and intake efficiency, such that the mass-flow rate and intake efficiency cannot be simultaneously maximised \cite{Romano2015,Romano2016,Binder2016}.

In these previous studies, the simplification of the aerodynamic modelling provides an optimistic representation of the satellite drag. Surface area parallel to the nominal oncoming flow has generally been neglected, despite being known to contribute to the aerodynamic drag. If solar arrays are required to generate the required power for the spacecraft, this can result in a design loop where the area for generating the required power also influences the propulsion system requirements (thrust, efficiency, mass-flow rate, and associated power). Furthermore, the variation of the oncoming flow direction and satellite attitude has generally not been considered in these studies. In inclined orbits the direction of the oncoming flow vector varies periodically during the orbit period due to the direction of the atmospheric co-rotation with the Earth. The contribution of thermospheric winds also provides further and less predictable variations in the true flow vector. Thus, unless the spacecraft can always align itself precisely with the true oncoming flow vector, the drag generated will generally be greater than in an idealised attitude. This is further complicated if off-axis pointing is required to perform mission operations, tracking of the sun vector is required by external solar arrays, or if moving aerodynamic control surfaces are considered.

\section{System Modelling Formulation}
This work builds on the foundations of system modelling for conventional satellites by incorporating the emerging technologies that are being developed to enable sustained operations VLEO. Through design- and trade-space exploration, the benefits provided by the emerging technologies can also be analysed, minimum requirements identified, and their effect on the system-level spacecraft design observed. In comparison to previous studies for satellites operating in reduced altitude, this work also seeks to provide a generalised and flexible approach to the system modelling for these technologies that can be applied to the development and study of a wide range of system concepts. 

A system modelling framework for VLEO satellite design is first needed that can be integrated with new models that represent the critical design parameters of the novel technologies. The conceptual design of VLEO satellites can then be explored for different mission profiles, applications, and assumptions of technology development. These concepts can subsequently be investigated in cooperation with market analysis, stakeholder identification, and business modelling work to establish those that provide promising economic viability. These results can also be used to influence the development of the novel technologies for use in future VLEO platforms.

\subsection{Multidisciplinary Analysis Framework}
\begin{figure*}
	\centering
	\caption{XDSM representation of the multidisciplinary analysis (MDA) framework for VLEO satellite design.} \label{F:DesignFramework}
	\begin{subfigure}[b]{\textwidth}
	\centering
	\includegraphics[width=180mm]{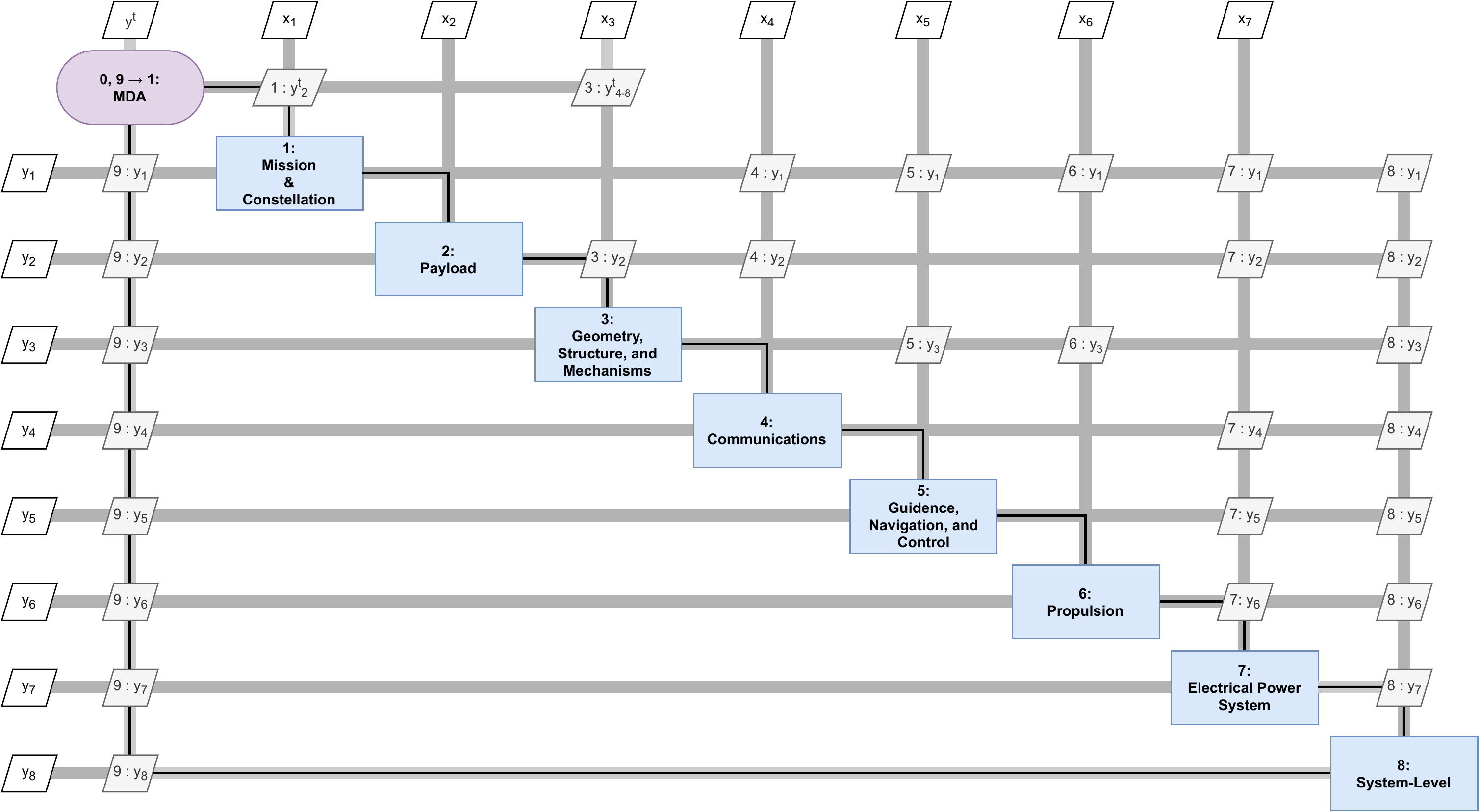}
	\label{F:MDA}
	\end{subfigure}
\newline
	\begin{subfigure}[t]{0.5\linewidth}
	\label{T:MDAInputs}
	\centering
	\begin{minipage}{\textwidth}
	\renewcommand*\footnoterule{}
	\renewcommand{\thempfootnote}{\arabic{mpfootnote}}
	\footnotetext[1]{Defined if ABEP is specified for design.}
	\begin{tabular}{m{0.05\linewidth} >{\raggedright\arraybackslash}m{0.85\linewidth}}
	\toprule
	\multicolumn{2}{c}{\bf{Input Variables}} \\
	\midrule
	$x_1$ & Orbit Type, Lifetime \\
	\cmidrule{2-2}
	$x_2$ & Resolution(s), Wavelength, Pixel Size, Quality Factor, Quantization, Payload Duty Cycle, Compression Ratio, Off-Nadir Pointing, Boresight Angle \\
	\cmidrule{2-2}
	$x_3$ & Material Properties, Safety Factors, Vibration Profile, Load Factors \\
	\cmidrule{2-2}
	$x_4$ & Architecture, Minimum Elevation, Latitude, Frequency, SNR, Downlink Rate, Bandwidth \\
	\cmidrule{2-2}
	$x_5$ & Pointing Accuracy, Pointing Knowledge, Slew Rate, Slew Time \\
	\cmidrule{2-2}
	$x_6$ & Propulsion System Type, Thruster Mode, Thruster Efficiency, Specific Impulse, Intake Area Ratio\footnotemark[1], Intake Duct Ratio\footnotemark[1], Intake Back Ratio\footnotemark[1] \\
	\cmidrule{2-2}
	$x_7$ & Array Efficiency, Array Specific Performance, Eclipse Efficiency, Sunlit Efficiency, Depth of Discharge, Specific Energy Density \\
	\bottomrule
	\end{tabular}
	\end{minipage}
	\end{subfigure}
\hfill
	\begin{subfigure}[t]{0.45\linewidth}
	\label{T:MDAOutputs}
	\centering
	\begin{tabular}{m{0.05\linewidth} >{\raggedright\arraybackslash}m{0.85\linewidth}}
	\toprule
	\multicolumn{2}{c}{\bf{Output Parameters}} \\
	\midrule
	$y_1$ & Altitude, Revisit, Eclipse Period \\
	\cmidrule{2-2}
	$y_2$ & Payload dimensions, Payload Power, Payload Mass, Downlink Rate \\
	\cmidrule{2-2}
	$y_3$ & Moments of Inertia, Centre of Mass, Aerodynamic Reference Area, SRP Reference Area, Drag Coefficient, Structure Mass \\
	\cmidrule{2-2}
	$y_4$ & Communications Mass, Communications Power, Antenna Size \\
	\cmidrule{2-2}
	$y_5$ & GNC Power, GNC Mass \\
	\cmidrule{2-2}
	$y_6$ & Thruster Power, Thruster Mass, Intake Size/Mass\footnotemark[2] \\
	\cmidrule{2-2}
	$y_7$ & Solar Array Area/Mass, Battery Mass \\
	\cmidrule{2-2}
	$y_8$ & Total System Mass, Total System Power, Total System Cost \\
	\bottomrule
	\end{tabular}
	\end{subfigure}
	\end{figure*}

The developed framework is shown in \cref{F:DesignFramework} in the form of an extended design structure matrix (XDSM) \citet{Lambe2012}. A sequential Gauss-Seidel multidisciplinary analysis (MDA) process is implemented to resolve co-dependencies that exist between some of the individual analysis functions. In this process each disciplinary analysis block is executed in sequence using the external design variables (inputs) and the most recent information from the preceding modules. This process is iterated until the state variables have converged. In the XDSM diagram (\cref{F:DesignFramework}), the thick grey lines represent possible data connections whilst the thinner black lines represent the process connections. The MDA block ($i=0$) distributes the information to the different subsystem analysis functions ($i=1-7$) and the system-level analysis ($i=8$) in the iterative scheme and checks for convergence of the process. The set of input design variables for each analysis is given by $x_i$ and output parameters from each analysis function given by $y_i$, both reported in \cref{F:DesignFramework}. Parameters identified by $y^t_i$ are target coupling variables (i.e. analysis inputs that are derived from prior outputs of the iterative scheme). 

By varying the input parameters to this MDA process and the internal analyses, parametric design space exploration can be performed, allowing investigation and identification of the trade-offs associated with different system requirements, environmental conditions, design decisions, or technological capabilities and performance. Whilst not implemented in the current study, this approach has also been developed to be compatible with multi-disciplinary optimisation processes, including both single- and multi-objective methods, enabling directed or ``intelligent'' search for an optimal design or sets of optimal designs based on defined ranges and/or constraints for different input variables.

For VLEO platforms, new analysis components are principally required within i) the propulsion module to incorporate the novel ABEP concepts, ii) the geometry, structure, and mechanisms module to account for the aerodynamic performance of the external spacecraft configuration, and iii) the guidance, navigation, and control module if aerodynamic control manoeuvres are to be considered.

This paper aims to exploring the platform trade-offs associated with the development and integration of new technologies that can enable sustained operations at lower altitude. As such, the developed analyses are focused on the use of circular or near-circular orbits in VLEO. The design of multi-satellite systems or constellation configurations to achieve a given performance with respect to revisit time or coverage is also not be treated in the present work.

Given the benefits of operating in VLEO for EO missions, the design of the payload, attitude determination and control system (ADCS), and communications subsystem are of interest and require some adaptation for platform design in lower altitude orbits. The power system is also specifically considered as it provides critical infrastructure for the spacecraft and is affected considerably by the propulsion specification and aerodynamic design. These fundamental relationships and system models are detailed in \cref{S:Models}. However, for much of the remaining systems design, relationships and methods for the conceptual and preliminary design of conventional satellites can be adopted from existing literature. A summary of the resources used in this work is provided in \cref{T:ModelsLit}.

\begin{table*}[tb]
	\caption{Summary of conventional system modelling resources.} \label{T:ModelsLit}
	\centering
	\begin{tabular}{m{0.35\linewidth}m{0.25\linewidth}m{0.1\linewidth}}
	\toprule
	\bf{Module} & \bf{Calculation} & \bf{References}\\
	\midrule
	\multirow{3}{*}{1: Mission and constellation} & Coverage & \cite{Wertz2011} \\
	& Revisit & \cite{Crisp2018} \\ 
	& Eclipse & \cite{Vallado2013} \\ [0.5em]
	2: Payload & Field of view & \cite{Wertz2011,Vallado2013}  \\ [0.5em]
	3: Geometry, structure and mechanisms & Structural mass & \cite{Wertz2011} \\ [0.5em]
	4: Communications & Link budget & \cite{Wertz2011} \\	[0.5em]
	\multirow{2}{*}{5: Guidance, navigation and control} & Disturbance torques & \cite{Wertz2011} \\
	& Actuator sizing & \cite{Wertz2011,Votel2012} \\ [0.5em]
	6: Propulsion & Propellant tank sizing & \cite{Humble1995,Chiasson2012} \\ [0.5em]
	7: Power & Battery and array sizing & \cite{Wertz2011} \\ [0.5em]
	8: System & Cost-estimating relationships & \cite{Wertz2011,Mahr2020} \\
	\bottomrule
	\end{tabular}
	\end{table*}

\subsection{VLEO Spacecraft Concepts and Configurations}
The basic configuration and geometric design of a spacecraft is principally driven by the requirements of the payload and the mission it is designed to perform. For EO spacecraft this typically involves specific pointing requirements and may include stability or agility requirements. For VLEO spacecraft the stability in the presence of aerodynamic forces and torques and the option of performing drag compensation are also critical considerations at the conceptual design level.

The concept matrix given in \cref{T:ConceptMatrix} can be used to generate different concepts for VLEO platforms. Combinations of the different options in each row will produce conceptual spacecraft designs, some of which will be considerably different to each other. Other combinations may be infeasible or irrational due to incompatibilities between the selected options, for example a tumbling spacecraft with a continuously-thrusting propulsion system for drag compensation.

\begin{table*}[tb]
	\caption{VLEO satellite concept development matrix.} \label{T:ConceptMatrix}
	\centering
	\newcolumntype{C}[0]{>{\centering\arraybackslash}p{0.03\linewidth}}
	\begin{tabular}{m{0.2\linewidth}CCCCCCCCCCCCC}
	\toprule
	& \multicolumn{12}{c}{\bf{Options}} \\
	\midrule
	\bf{Propulsion} & \multicolumn{3}{c|}{Chemical} & \multicolumn{3}{c|}{Electric} & \multicolumn{3}{c|}{ABEP} & \multicolumn{3}{c}{None} \\
	\midrule
	\bf{Drag Compensation} & \multicolumn{4}{c|}{Continuous} & \multicolumn{4}{c|}{Periodic} & \multicolumn{4}{c}{None} \\
	\midrule
	\bf{Material GSI} & \multicolumn{6}{c|}{Current (Diffuse)} & \multicolumn{6}{c}{Novel (Quasi-specular)}\\
	\midrule
	\bf{Stability} & \multicolumn{3}{c|}{Three-Axis} & \multicolumn{3}{c|}{Aerostable} & \multicolumn{3}{c|}{Spinning} & \multicolumn{3}{c}{Tumbling} \\
	\midrule
	\bf{Control} & \multicolumn{4}{c|}{Internal Actuators} & \multicolumn{4}{c|}{Control Surfaces} & \multicolumn{4}{c}{Mixed} \\
	\midrule
	\bf{Payload} & \multicolumn{4}{c|}{Static} & \multicolumn{4}{c|}{Steerable} & \multicolumn{4}{c}{Hybrid} \\
	\midrule
	\bf{End of Life} & \multicolumn{6}{c|}{None} & \multicolumn{6}{c}{Additional/Existing Device} \\
	\bottomrule
	& & & & & & & & & & & & \\ [-1em] 
	\end{tabular}
	\end{table*}

In VLEO a propulsion system may be required to provide a useful lifetime. If the aerodynamic drag is to be compensated for, thrust should be provided opposite to the velocity vector with associated constraints on the platform configuration. Furthermore, if an ABEP system is to be used to enable sustained operations, the intake should face the oncoming flow direction to provide maximum efficiency.

A surface area for power raising using solar cells is necessary to provide sustained operational performance. Body-mounted solar cells are preferred to avoid increment in the experienced drag but deployed solar arrays may be required to meet increased power demands, particularly if EP or ABEP systems are to be used. Furthermore, the orientation of the sun with respect to the orbit and satellite may mean that body mounted solar cells may not all be simultaneously illuminated or operating at their full efficiency. The orientation of non-body-mounted solar arrays therefore requires careful consideration to balance their power generation capability with their contribution to aerodynamic drag.

The centre of mass of the spacecraft and relative location of any deployable or extended surfaces also requires careful consideration to ensure stability and control of the platform, particularly in the presence of potentially disturbing aerodynamic and solar radiation perturbations. In VLEO, the geometric design of the satellite can be used to provide aerostability. However, this may simultaneously reduce the agility of the satellite and the ability to point away from the oncoming flow direction, for example if payload pointing is required away from the nominal aerostable attitude. Alternative approaches to payload design may be able to compensate for such designs, for example using steerable optical assemblies that reduce the need for attitude control manoeuvres to perform off-axis target acquisition.

A key aim of most geometric designs for VLEO is to reduce the aerodynamic drag experienced by the satellite. Slender spacecraft configurations are therefore recommended as the projected area (i.e. cross-sectional to the flow) is minimised and the remaining area is parallel to or shielded from the flow, reducing drag. Such configurations are also suited to providing natural aerostability. However, alternative configurations can also be considered, including those that use attitude actuators to provide conventional three-axis control, make use of neutral stability providing high agility, or are equipped with external control surfaces that can enable aerodynamic orbit and attitude control.

The use of surfaces that are angled with respect to the nominal oncoming flow may be used to mitigate the effects of aerodynamic forces and reduce the experienced drag \cite{Walsh2020}. The magnitude of this potential reduction is dependent on the angle with respect to the flow and the GSI properties of the materials on the external spacecraft surfaces. The useful internal volume of such geometric configurations, the corresponding effect on stability, also and whether the forward and aft facing surfaces are required for other purposes (e.g. thrusters and atmospheric intakes) also requires consideration.

In VLEO, end of life deorbit is naturally enabled and quickly realised as a result of the increased aerodynamic drag. However, if very low-drag platforms are realised using novel aerodynamic materials, active methods of deorbit may still be needed to avoid unnecessary occupation of orbits by non-operational spacecraft that could contribute to an increased risk of collision. Targeted re-entry may also be necessary for very large VLEO spacecraft that are challenging to ``design-for-demise'' and may not completely disintegrate during re-entry, presenting a risk of casualty or damage at ground level.

\subsection{Conceptual VLEO Satellite Configuration}
\begin{figure}[tb]
	\centering
	\includegraphics[width=90mm]{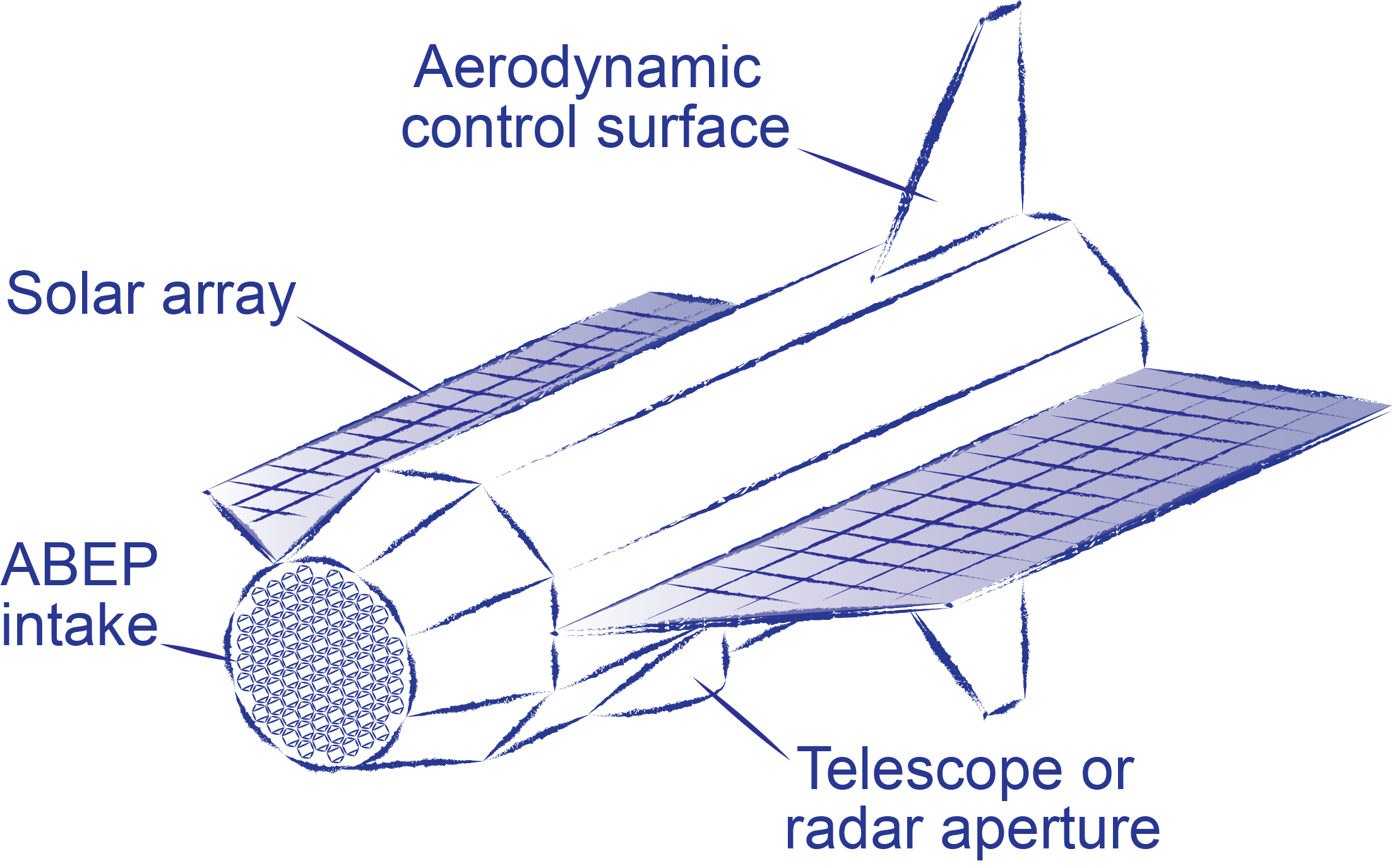}
	\caption{Representative concept geometry for a slender ``arrow-like'' spacecraft.}
	\label{F:Arrow}
\end{figure}

Conceptual configurations of satellite platforms can be useful in informing the development and implementation of the new system models and analyses for VLEO technologies. A slender ``arrow-like'' configuration, shown in \cref{F:Arrow}, has been adopted in this work. An analogous configuration was used operationally in VLEO previously by the GOCE spacecraft, albeit without the ABEP intake. The proposed ``Skimsat'' platform also exhibits features of this basic configuration \cite{Bacon2017}.

This geometry is designed to be nominally aligned lengthwise with the velocity vector and therefore close to the direction of the oncoming flow. As a result, this configuration has a relatively low cross-sectional area with the aim of reducing the experienced drag. Given this nominal attitude, this configuration is also generally compatible with both EP and ABEP systems as the thruster can be appropriately aligned with the oncoming flow direction. The front (ram) facing surfaces of the spacecraft can also be used to accommodate an atmospheric intake, if required, or tapered to further reduce the aerodynamic drag if permitted by the material GSIs.

Extended surfaces, oriented parallel to the length of the satellite body, can be utilised for power raising purposes whilst minimising the increment in drag force. When located behind the centre of gravity, these surfaces can also contribute to natural aerodynamic stabilisation of the spacecraft. Active control surfaces may also be considered for the purpose of aerodynamics-based attitude and orbit control manoeuvres.

To achieve the best propulsive efficiency, pointing of the spacecraft away from the velocity vector in pitch and yaw should be avoided. Payloads that can be operated in a fixed orientation are therefore most suitable, though pointing in roll can be permitted without increasing drag or compromising the propulsive performance. Alternatively, the use of steerable optical or radar payloads can be considered to provide more flexible observational performance. A similar compromise for any extended solar arrays exists between the increase in drag and alignment with the solar vector for maximum power-raising performance. This is principally dependent on the orbit geometry and relative angle of the sun with respect to the spacecraft.

\subsection{System Models for VLEO Spacecraft} \label{S:Models}
System models with a significant dependence on orbital altitude or that are of particular interest in the VLEO altitude range are outlined in the following subsections. Models for the state-of-the-art and emerging technologies that may contribute to the development of novel VLEO platforms are also described.

\subsubsection{EO Payload Sizing}
The basic payload design relationships for EO platforms operating in VLEO do not differ substantially to those of conventional EO satellites in principle. However, the configuration, orientation, and compactness of the payload with respect to the spacecraft geometry may be a limiting factor on the aerodynamic performance. For a typical optical payload, the minimum aperture diameter $D$ to provide a given diffraction limited ground resolution distance $\mathrm{GRD}$ can be approximated by considering the range $R$ to the target and the wavelength $\lambda$ of interest.
\begin{equation}
D \approx 1.22 \frac{\lambda R}{\mathrm{GRD}}
\end{equation}

The required focal length $f$ can similarly be calculated from the altitude $ h_\phi$ and the physical size of the detector elements $x$. 
\begin{equation}
f = \frac{R^2 x}{\mathrm{GRD} \, h_\phi}
\end{equation}

A range of different configurations for optical assemblies can be considered that have various advantages and characteristics for high resolution imaging, e.g. number of mirrors, mechanical tolerances, compactness, and stability. Examples include Korsch, three mirror anastigmatic (TMA), and catadioptric designs \cite{Costes2012,Metwally2020}. To capture the compactness of these different assembly types, a geometric reduction factor can be used, avoiding the need to develop a complete telescope design to estimate the axial dimension and volume of the payload from the focal length and diameter. The mass of an optical payload can subsequently be estimated heuristically by using sizing relationships (i.e. mass density) informed by existing satellite imaging systems. The corresponding power required by such a payload can either be estimated similarly from existing systems using a power density relationship or can be set a priori based on knowledge of the required performance of the payload (e.g. spectral capability or on-board data processing and storage).

For a SAR instrument the minimum required antenna area and maximum antenna length can be calculated principally from resolution requirements and ambiguity constraints \cite{Tomiyasu1978,Cutrona1990}. The minimum width of the antenna can subsequently be determined, though a larger area and width may be desired to provide a design margin. The average power required by the SAR antenna $P_\mathrm{SAR}$ can be calculated using \cref{E:SAR} from the range resolution $\delta_R$, range $R$, antenna area $A_\mathrm{ant}$, and sensitivity $\sigma^0_\mathrm{NE}$ given as the noise-equivalent sigma-zero (NESZ):
\begin{equation} \label{E:SAR}
P_\mathrm{SAR} = \frac{8 \pi k_B T_r \overline{\mathit{NF}} R^3 V \lambda l_s} {A_\mathrm{ant}^2 \eta_\mathrm{ant}^2 \delta_R \sigma^0_\mathrm{NE}}
\end{equation}
where $k_B$ is Boltzmann's constant and $V$ is the satellite orbital velocity. In the early design phase assumption of the antenna efficiency $\eta_\mathrm{ant}$, noise factor $\overline{\mathit{NF}}$, receiver noise temperature $T_r$, and system losses $l_s$ are also generally necessary.

The mass of the SAR instrument can be approximated using a value of area density determined from similar instruments. However, the instrument design and mass are related to a number of other performance factors, for example number of image acquisition modes and need for polarisation diversity \cite{Freeman2018}. The use of a given antenna area density will therefore only reflect a small range of the SAR instrument configurations or antenna types and will be specific to a given set of mission requirements.

\subsubsection{Electrical Power System}
Electrical power system (EPS) architecture for VLEO satellites is likely to be similar to that of conventional orbiting platforms, using solar arrays as the primary power source with supporting batteries to enable operations during eclipse periods. However, for VLEO satellites with ABEP or EP for drag compensation or mitigation, power requirements are likely to be significantly increased. Furthermore, eclipse conditions in VLEO also increase slightly in duration with reducing altitude. Even in dawn-dusk or similar SSOs, often considered due to their favourable on-orbit illumination conditions, the length of the eclipse seasons also increases with reducing altitude \cite{Fearn2005}.

For solar-based power the implication on spacecraft aerodynamics (principally drag) and stability due to the presence of extensive solar arrays must be carefully considered and captured in the aerodynamic models. If the solar arrays are expected to remain aligned parallel with the direction of the flow to minimise drag, the loss in power-raising performance associated non-alignment of the arrays with the sun vector must also be incorporated into the power system and sizing calculations to ensure that sufficient power is available throughout the orbit. Dawn-dusk SSOs allow flow-aligned solar arrays to remain approximately normal to the solar vector, minimising the cosine loss and maximising power-raising performance. For generalised inclined orbits, the worst-case condition will exist when the beta angle is \ang{0}. For fixed solar arrays oriented in the local horizontal plane under this condition, where $\theta$ is the angle between the solar array normal and the solar vector, the cosine loss can be approximated over the period of an orbit.
\begin{equation}
\frac{1}{2\pi}\int_{-\pi}^{\pi} |\cos{\theta}|\,d\theta = \frac{2}{\pi} \approx 0.637
\end{equation}

Given the challenges and high requirements for EPS design in VLEO, the system performance is critical. High efficiency solar cells help to reduce the solar array area and therefore reduce contributions to aerodynamic drag and the thrust requirement. High specific power of solar arrays and increased energy density of batteries will also contribute to a reduction in system mass.

\subsubsection{Aerodynamic Performance}
Estimation of the satellite aerodynamic performance is critical in VLEO as it is required to either determine the lifetime of the spacecraft before deorbit, the propulsive and power requirements to mitigate or compensate the drag and extend or sustain the mission, or to determine the capability of the satellite to perform aerodynamic attitude and orbit control manoeuvres. 

Aerodynamic forces $\bm{F_A}$ in three dimensions can be calculated using \cref{E:Force} from the atmospheric density $\rho$, relative velocity $\bm{v_{\mathrm{rel}}}$ (with respect to the oncoming flow), reference area $A_{\mathrm{ref}}$, and set of aerodynamic force coefficients $\bm{C_F}$ that are defined in axes that correspond to the output force. 

\begin{equation} \label{E:Force}
\bm{F_A} = \frac{1}{2} \rho {\|\bm{v_{\mathrm{rel}}}\|}^2 A_{\mathrm{ref}} \bm{C_F}
\end{equation}

The drag force $F_D$ is considered as the component that is aligned with the orbital velocity vector of the satellite and is therefore usually closely aligned with the direction of the oncoming flow. Mutually perpendicular forces (herein termed lift) can also be considered. However, the predominant GSI characteristics of typical spacecraft materials in VLEO are diffuse and the lift forces generated are typically an order of magnitude smaller than the drag force. If specularly reflecting GSI performance is considered the magnitude of these forces can increase significantly depending on the angle of the surfaces with respect to the flow.

Aerodynamic torques $\bm{T_A}$ can also be produced by interaction of the satellite body with the atmospheric flow. These aerodynamic torques can be calculated from the aerodynamic force and associated moment arms $\bm{\ell}$ or in a similar manner to \cref{E:Force} by considering the set of aerodynamic moment coefficients $\bm{C_M}$ of the satellite geometry and a reference length $L_{\mathrm{ref}}$.
\begin{equation} \label{E:Torque}
\bm{T_A} =  \bm{\ell} \times \bm{F_A} = \frac{1}{2} {\|\bm{v_{\mathrm{rel}}}\|}^2 A_{\mathrm{ref}} L_{\mathrm{ref}} \bm{C_M}
\end{equation}

The sets of aerodynamic force and moment coefficients of a body in VLEO are dependent on the external geometry of the spacecraft, the orientation of the spacecraft with respect to the flow, the oncoming flow conditions, and the GSI characteristics between the surface materials and the atmospheric flow. In this study Maxwell's model \cite{March1878} has been selected to explore the effect of different materials on the aerodynamic performance in VLEO. This GSI model combines diffuse and specular particle interactions using an accommodation coefficient ($\alpha$) that defines the proportion of particles that are fully accommodated and re-emitted thermally and diffusely whilst the remaining proportion are reflected elastically and specularly. However, whilst capable of describing the effect of some specular reflection qualities, Maxwell's model is not able to adequately describe the more complex petal-like shaped distribution of materials with a predominant specular or quasi-specular component that has been observed experimentally \cite{Livadiotti2020}. 

A range of alternative GSI models have been developed with different assumptions, limitations, and level of complexity. However, significant uncertainty remains associated with the aerodynamic performance of even commonly used materials in the VLEO environment, not to mention the possible characteristics of novel atomic oxygen resistant materials that are currently being sought to reduce drag in VLEO. As knowledge of these materials improves and their GSI properties become more fully characterised, models that more appropriately reflect these behaviours can be incorporated into future studies.

The density at a given orbit altitude can be estimated using an atmosphere model. Different atmosphere models of varying complexity and fidelity can be used with possible inputs dependent on geographic position, time, and space weather \cite{AIAA2010}. However, for conceptual design purposes the fine detail of the atmospheric density is not necessary and uncertainties in estimating the future space weather indices are considerable. A characteristically high or averaged value of the density at a given altitude can therefore be used as a conservative estimate. The NRLMSISE-00 \cite{Picone2002} model is commonly used and openly available and has therefore been used in this study in conjunction with input parameters as recommended in ISO 14222:2013 \cite{ISO2013}.

For simple spacecraft configurations, the overall aerodynamic drag coefficient and associated reference area can be calculated by the panel method, i.e. by summing the contribution of a set of simple flat panels that provide a basic representation of the spacecraft geometry.
\begin{equation} \label{E:PanelMethod}
\{C_D A_{\mathrm{ref}}\}_{T} = \sum_{i=1}^n C_{D,i} A_i
\end{equation}

This basic method can be extended and applied to CAD or mesh-based geometries for more complex satellite representations. The panel method tool ADBSat \cite{Mostaza-Prieto2017a,Sinpetru2021a} has been used in this study and has been shown to demonstrate good agreement with more complex simulations (e.g. direct simulation Monte Carlo) whilst remaining easy to implement for use in preliminary design studies. Furthermore, this tool can be adapted for use with different GSI models under the general assumptions that the geometry used is generally convex (i.e. does not have surfaces that can experience secondary reflections) and the flow is strictly free-molecular.

\subsubsection{Propulsion}
For VLEO spacecraft, the principal function of a propulsion system is to perform either mitigation or complete compensation of the aerodynamic drag to extend or sustain operations. However, if appropriately specified and designed a propulsion system may also be used to perform orbit-raising or lowering procedures, constellation maintenance, out-of-plane manoeuvres, contribute to attitude control, or support a controlled/targeted re-entry as required by different mission concepts.

For both conventional EP systems (for example demonstrated by GOCE) and novel ABEP systems in VLEO, the thrust required for continuous drag compensation must be equal the drag force that will be experienced over the expected lifetime in orbit. However, alternative thrusting strategies can be considered that may reduce the power required during eclipse, for example periodic thruster operation either only during sunlit conditions or at another chosen ratio \cite{DiCara2007}. Under such strategies, the thruster must provide sufficient thrust to ensure that the effect of the drag over an orbit period (or similar time scale) is compensated for and a mean altitude is maintained. The required thrust force $F_T$ can be approximated by considering the accumulated drag force over a given orbit divided by the period of the orbit that the thruster will be operated for.
\begin{equation}
F_T \approx \frac{F_D \, t_{\mathrm{period}}}{t_{\mathrm{thrust}}}
\end{equation}

Periodic thrusting strategies may also be considered if off-axis pointing performance is required in either the pitch (or less commonly yaw) axis, causing a significant misalignment of the spacecraft from the desired thrust vector. Such strategies would allow periods of agile EO operation whilst compatibility with a drag-compensating propulsion system is maintained. However, it is likely that this imaging would take place during sunlit conditions and therefore conflicts with the optimal drag-compensation phase and power availability. The impact on the overall system performance will therefore be of greater consequence.

\paragraph{Conventional Electric Propulsion} \hfill \\
A conventional EP system comprises primarily of a thruster, power supply hardware, a propellant feeding system, and the required propellant that is typically stored within a pressurised tank. For spacecraft operating in VLEO, a high specific impulse is typically desired to minimise the mass of propellant required.

Under the assumption that the propulsion system operates continuously to compensate the aerodynamic drag force most efficiently for the total length of the mission $t_L$, the mass of propellant can be estimated from the specific impulse $I_{sp}$. A mass usage efficiency of the thruster $\eta_m$ can also be introduced.
\begin{equation} \label{E:PropMass}
m_p = \frac{F_D}{I_{sp} g_0 \eta_m} t_L
\end{equation}

The power required by the thruster can be calculated from the required thrust $F_T$, propellant mass flow rate $\dot{m}_p$, and total thruster efficiency ($\eta_T = \gamma^2 \eta_m \eta_e$), where $\eta_e$ is the electrical efficiency of the thruster and $\gamma$ is a thrust correction factor (accounting for multiply charged ions and beam divergence) \cite{Goebel2008}.
\begin{equation}
P_t = \frac{F_T^2}{2 \dot{m}_p \eta_T}
\end{equation}

For EP, a trade-off between thrust, input power, and specific impulse results. For a given thrust requirement, increasing $I_{sp}$ can reduce the necessary propellant mass (see \cref{E:PropMass}), but requires a corresponding increase in input power.
\begin{equation}
F_T = \frac{2 \eta_T P_T}{g_0 I_{sp}}
\end{equation}

Given the required thrust magnitude, mass of propellant, and power consumption, the basic propulsion system mass can be estimated using simple sizing relationships and historical database of actual system parameters and simple regression methods \cite{Humble1995,Chiasson2012}. Typical efficiencies for Hall Electric Thrusters (HETs) and Gridded Ion Engines (GIEs) range from \numrange{0.20}{0.90}, whilst specific impulses can range from \SIrange{600}{10000}{\second} \cite{Holste2020}.

\paragraph{Atmosphere-Breathing Electric Propulsion (ABEP)} \hfill \\
In contrast to conventional EP, on-board propellant is not required for an ABEP system as the reaction mass is collected by the spacecraft from the surrounding atmosphere using an intake. For conceptual and preliminary design purposes, the ABEP system can be considered simply as two discrete components, the intake and the thruster. 

Given the magnitude of the required thrust $F_T$, the power required by the ABEP thruster $P_t$ can be calculated from the external mass flow rate into the intake ($\dot{m}_c = \rho A_c v$), the intake collection efficiency $\eta_c$, and the thruster efficiency $\eta_t$.
\begin{equation}
P_t = \frac{F_T^2}{2 \dot{m}_c \eta_c \eta_T}
\end{equation}

Critical considerations for ABEP thruster development include compatibility and performance under the conditions of changing atmospheric density and composition, for example with the solar cycle or smaller short-term variations. The technical challenges of thruster design for operation with variable and mixed atmospheric propellant must also be considered. Component degradation (e.g. electrode erosion principally by atomic oxygen) must also be avoided to ensure long-term operation and has resulted in increasing interest in electrodeless and contactless thruster designs that utilise radio-frequency (RF) plasma discharges. Owing to the relative immaturity of these technologies, the efficiency of RF helicon thrusters is currently lower than that of established EP types, and has currently demonstrated maximum thruster efficiencies of \numrange{0.2}{0.25} \cite{Hohman2012,Takahashi2019}. Given the novelty of ABEP thrusters, relationships between their mass for different power or thrust requirements cannot yet be developed. However, given the similarity in technology, an analogous relationship to RF ion thrusters can be considered to provide an estimate of the mass of a notional ABEP thruster. 

A number of designs for rarefied intakes have been presented in the literature \cite{Singh2015}. However, many of these designs are given for a specific set of mission or platform design parameters and cannot be simply scaled for alternative scenarios. Studies of parametrised ABEP intakes allow scaling based on frontal area and thruster specification using a set of geometric design factors (e.g. inlet-outlet area ratio and length-radius ratios for internal features). The collection efficiency $\eta_c$ can be calculated using a balance model and transmittance probabilities for rarefied flows, but such methods can only be applied to diffusely re-emitting surfaces \cite{Romano2015,Romano2016,Binder2016}. These analytical models have been compared to simulations using direct simulation Monte Carlo and particle-in-cell methods. Deviations were found the range of \SIrange{5}{8}{\%}, demonstrating good agreement and their suitability particularly for preliminary design studies. The maximum efficiency of these diffuse intakes has been shown to range up to approximately \num{0.60} \cite{Binder2016}. If specular reflecting materials can be identified and integrated for use in the orbital environment then alternative intake designs can be developed, and higher intake efficiencies (up to \num{0.70} for hybrid designs and \num{0.94} for pure-specular reflections) can be theoretically achieved \cite{Romano2021}. However, flexible design and optimisation methods for such intakes are currently lacking and computationally expensive Monte Carlo simulations are required to establish their performance. For such intakes, the mass of the assembly can be estimated from the geometric properties, a structural wall thickness, and selected material properties. In the presence of significant uncertainty associated with the knowledge of material GSI performance and ABEP intake performance a priori values of intake efficiency can be used for simplicity and to explore the design space as this area of system design and modelling matures.

The use of ABEP can present a significant challenge in system design convergence and optimisation. The thruster power requirement is simultaneously dependent on the intake efficiency, thrust requirement, and available mass flow rate. Meanwhile, the intake efficiency is principally dependent on the inlet-outlet ratio that also affects the mass flow rate to the thruster. At low altitude, the power requirement needed for drag compensation increases due to the increased atmospheric density and therefore experienced drag. This increases the area required for power raising (assuming the use of solar arrays), which in turn simultaneously increases the drag and the power requirement. To reduce the power requirement, a larger mass flow to the thruster is desired, requiring either a more efficient intake or a larger intake area that would once again increase the drag and the power. A circular dependency that can cause non-convergence is therefore present in the design and a trade-off between these characteristics has to be matched to ensure sustained operation at lower altitude \cite{Romano2016}. At higher altitude where the thrust and mass flow rate requirements are lower, it may be advantageous to consider intake designs that only cover a fraction of the cross-sectional area, reducing the mass of the intake. Furthermore, if specular reflecting materials are identified, the remaining forward-facing surfaces can be angled with respect to the flow, reducing the overall drag of the satellite, and therefore reducing the power requirement of the propulsion system.

\subsubsection{System Cost}
Parametric cost models can be used to estimate the development and manufacturing costs for future space systems during the early design phases. These models typically utilise a set of cost-estimating relationships (CERs) that express system or subsystem costs as a function of different performance parameters. These CERs are generally developed from databases of historic satellites using regression-based analysis \cite{Wertz2011}. The Unmanned System Cost Model (USCM) and the Small Satellite Cost Model (SSCM) are publicly available and widely used. The NASA Instrument Cost Model (NICM) can also be used to provide estimations for specific payload costs \cite{Foreman2016}. The USCM and SSCM also provide estimation of model errors, allowing for some measure of the uncertainty in the cost estimation process. The CERs contained in the USCM8 \cite{Wertz2011}, SSCM19 \cite{Mahr2020}, and NICM Version IIIC \cite{Wertz2011} have been used in this work.

Whilst these models can provide some representation of cost for the system development and production, they currently lack proper compatibility with VLEO platform designs that may incorporate novel technologies, for example ABEP and aerodynamic surface coatings. Furthermore, given the low technology readiness level (TRL) of these technologies, there is currently insufficient data to produce new regressions that can provide reliable cost estimates. In the absence of suitable CERs, the cost of the ABEP system will be estimated analogously; the intake will be associated with the structural subsystem whilst the thruster and supporting systems are considered as components of a conventional propulsion system. At present, the cost-implications of aerodynamic surface coatings are largely unknown and will not be considered.

It should also be noted that each cost model and the associated CERs are limited in their range of application and inputs (e.g. total system and individual subsystem masses) as a result of the underlying data used in their development. For example, the SSCM does not provide a CER for propulsion systems for spacecraft with a wet mass below \SI{100}{\kilo\gram}. Discontinuities will therefore be present across the transitions between the different models.

When the development of multiple spacecraft is considered, for example in constellations, learning curves can be incorporated to account for the reduction in recurring production costs that can be achieved. Further economic considerations such as adjustments for inflation and amortisation for projects lasting multiple years can also be included but have been neglected at present.

Given the inherent cost model uncertainties and additional issues with cost modelling for VLEO systems described above, it is recommended that the output system costs presented in the following section are used only to identify rough system-level trends and to compare different designs in a relative and not absolute sense.

\section{System Case Studies}
Using the described system design framework and developed system models, it is possible to investigate the system-level design of VLEO spacecraft and the trade-offs associated with the emerging technologies of interest. Two case-studies will be considered; the first a very-high resolution (VHR) optical imaging satellite, and the second a synthetic aperture radar (SAR) satellite. Sets of input parameters are presented for each case to explore the different VLEO-specific technologies and to represent their near-term expected performance and longer-term speculative properties. In both case studies, the current state-of-practice and improvements to the supporting technologies will also be considered, notably solar array efficiency and specific power, and battery specific energy density.

\subsection{Optical Very High Resolution (VHR)}
The reduced altitude of VLEO lends itself to VHR imagery as the distance to the target is reduced and thus the payload requirements for a given resolution reduce. In combination with VHR, a high-performance EO satellite would require the high image stability and quality, precise geolocation, and flexible tasking and pointing performance. If deployed in a distributed system or constellation, daily or sub-daily revisit would likely be desired.

Existing systems in this segment include WorldView-1/2/3/4, Pleiades-1A/1B, and Kompsat-3/3A that serve principal application areas of defence monitoring and mission planning, construction and infrastructure monitoring, and urban development. A satellite with the aim of replacing such systems would require a GSD of \SI{0.3}{\meter} and operate for at least 5~years. The principal mission design parameters for a notional satellite of this class are given in \cref{T:VHRin}.

\begin{table}
	\caption{Optical VHR mission and system input parameters.} \label{T:VHRin}
	\centering
	\begin{tabular}{l r}
	\toprule
	\bf{Parameter} & \bf{Value} \\
	\midrule
	Design lifetime [\si{years}] & 5 \\
	$\mathrm{GRD}$ (nadir) [\si{\meter}] & 0.3 \\
	VHR EO Payload Density [\si{\kilo\gram\per\meter\cubed}] & 61.5 \\
	Payload Power [\si{\watt}] & 400 \\
	Orbit & SSO \\
	LTAN & 10:00h \\
	Maximum off-nadir pointing angle [\si{\degree}] & 30 \\
	Maximum slew rate [\si{\degree\per\second}] & 4.5\\
	\bottomrule
	\end{tabular}
	\end{table}

\begin{figure*}
\centering
	\includegraphics[width=180mm]{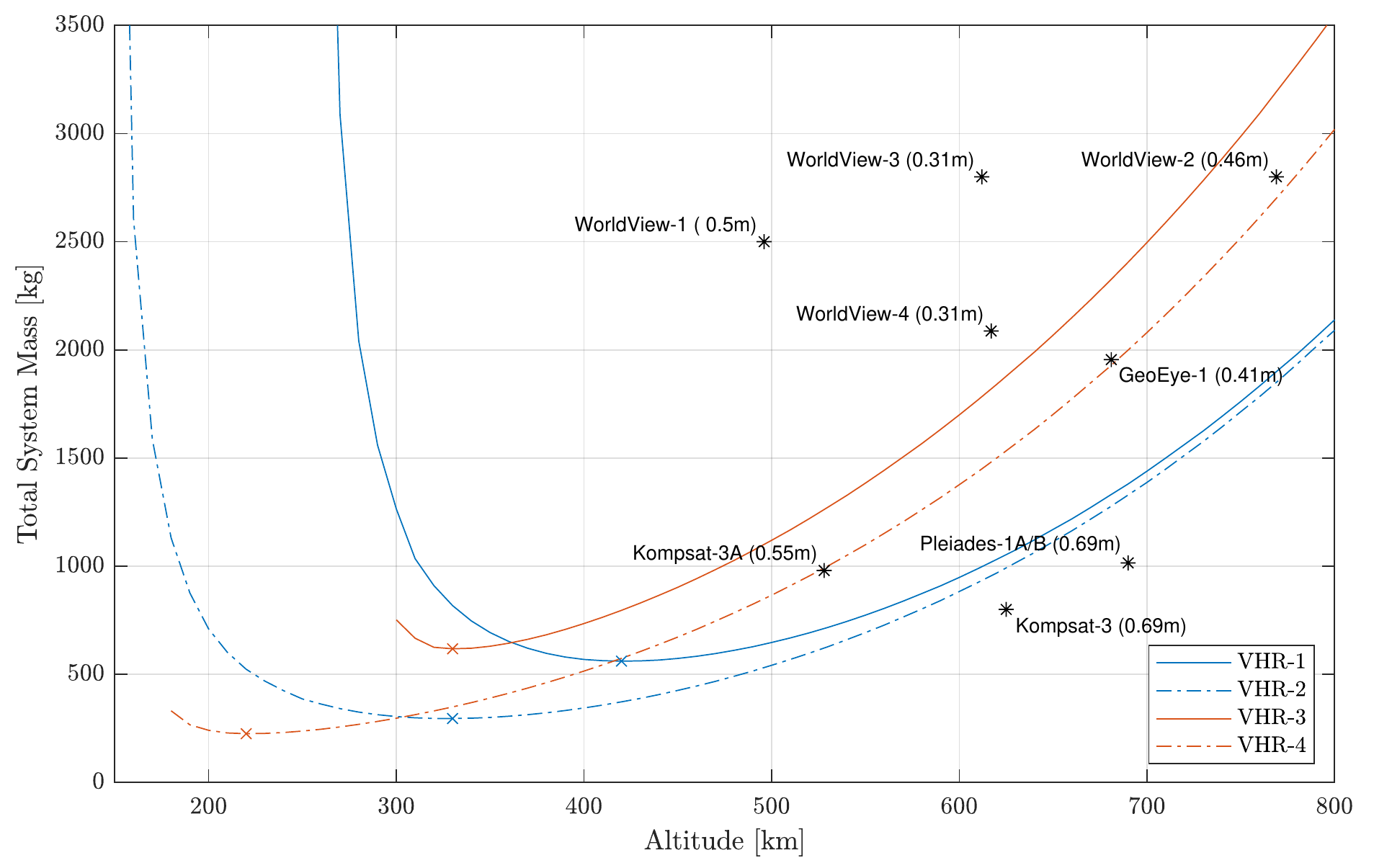}
	\caption{Variation of system mass with altitude for different optical VHR satellite designs (\SI{0.3}{\meter} GSD). The point of minimum mass for each case is marked with by an 'x'. Previous optical VHR satellite designs (with varying GSD noted) are shown for comparison.}
	\label{F:VHR_Altitude}
	\end{figure*}

\begin{figure*}[tb]
\centering
	\includegraphics[width=180mm]{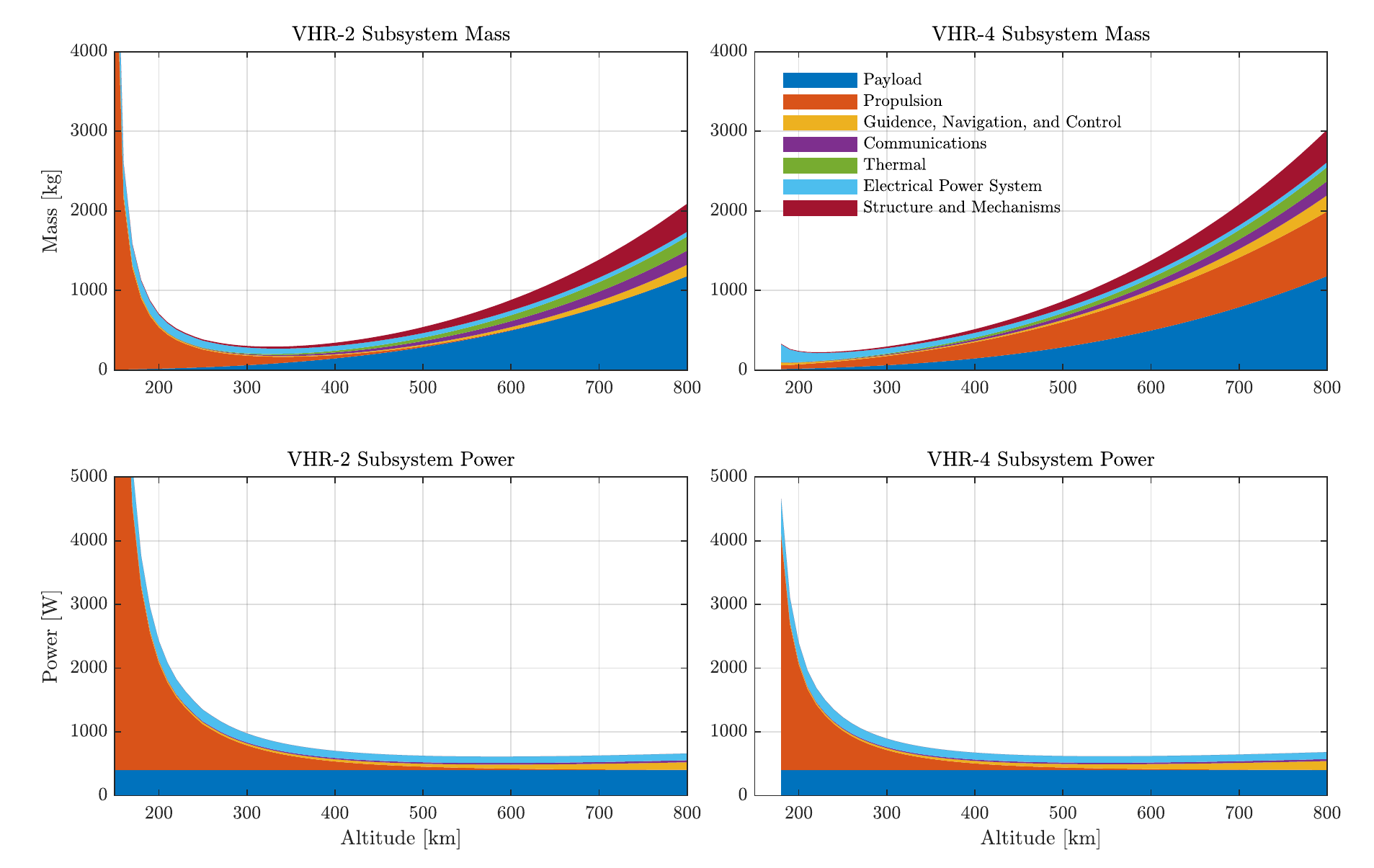}
	\caption{Variation of subsystem mass and power with altitude for different optical VHR satellite designs (VHR-2 and VHR-4).}
	\label{F:VHR_Subsystems}
	\end{figure*}

\begin{table*}
	\caption{Optical VHR satellite design input and output parameters.} \label{T:VHRout}
	\centering
	\begin{tabular}{l r r r r}
	\toprule
	\multirow{2}{*}{\bf{Parameter}} & \multicolumn{4}{c}{\bf{Case}} \\
	& VHR-1 & VHR-2 & VHR-3 & VHR-4 \\
	\midrule
	& \multicolumn{4}{c}{\bf{Inputs}} \\
	\midrule
	Propulsion Type 										& EP 	& EP 	& ABEP 	& ABEP 	\\
	Thruster Efficiency, $\eta_t$ [-] 						& 0.60 	& 0.80 	& 0.30 	& 0.60	\\
	Specific Impulse, $I_{sp}$ [\si{s}]						& 2500	& 4000	& -		& -		\\
	Intake Efficiency, $\eta_c$ [-]							& - 	& - 	& 0.40 	& 0.94 	\\
	Acc. Coefficient, $\alpha$ [-]							& 1.00 	& 0.50 	& 1.00 	& 0.50	\\
	Solar Array Efficiency [-]								& 0.30	& 0.40	& 0.30	& 0.40	\\
	Solar Array Specific Power [\si{\watt\per\kilo\gram}]	& 60	& 100	& 60	& 100	\\
	Battery Energy Density [\si{\watt\hour\per\kilo\gram}]	& 125	& 200	& 125	& 200	\\
	\midrule
	& \multicolumn{4}{c}{\bf{Outputs}} \\
	\midrule
	Altitude [\si{\kilo\meter}] 							& 420 	& 330 	& 330 	& 220 	\\
	Mass [\si{\kilo\gram}] 									& 561 	& 296 	& 618 	& 225 	\\
	Solar Array Area [\si{\meter\squared}]					& 14.3 	& 7.1 	& 25.2 	& 11.1	\\
	Aperture Diameter [\si{\meter}] 						& 0.85 	& 0.67 	& 0.67 	& 0.45 	\\
	Body Diameter [\si{\meter}] 							& 2.7	& 2.1 	& 2.1 	& 1.4	\\
	Unit Cost [\$M] 										& 347 	& 274 	& 292 	& 201 	\\
	\bottomrule
	\end{tabular}
	\end{table*}

Using the previously described system modelling framework, system designs can be generated for sets of different input options and parameters. Four different cases are defined (see \cref{T:VHRout}), illustrating the current and future expected performance of EP and ABEP enabled systems. For each case, the variation of modelled spacecraft mass with reducing altitude is shown in \cref{F:VHR_Altitude} alongside some recent examples of commercial VHR satellites. Whilst these existing systems are provided for high-level comparison, it should be noted that the systems modelling approach and inputs will not directly correspond to the specification, configuration, and performance of any of these spacecraft individually. Output design parameters of interest are reported in \cref{T:VHRout}, corresponding to the lowest mass system reported for each set of input parameters. The breakdown of mass and power by subsystem is also given in \cref{F:VHR_Subsystems} for cases VHR-2 and VHR-4, providing further insight into the system trade-offs associated with the selection and performance of the different technologies.

At an altitude of \SI{300}{\kilo\meter}, the optical payload to produce \SI{0.3}{\meter} resolution is estimated using the payload sizing method to have a mass of approximately \SI{60}{\kilo\gram}. In comparison, at \SI{600}{\kilo\meter} altitude such a payload would need to be approximately \SI{500}{\kilo\gram}, similar to that of the WorldView-4 satellite that has a \SI{550}{\kilo\gram} instrument and operates at an altitude of \SI{617}{\kilo\meter}.

When operating in VLEO, a satellite with a conventional EP system and assuming traditional materials (Case VHR-1) is shown to achieve the desired imaging performance at a lower total (wet) mass than existing systems, particularly those matching the high-resolution requirement. This result is consistent with the success of the GOCE spacecraft that had a payload mass of approximately \SI{200}{\kilo\gram} and was able to operate for a period of over \SI{4}{years} at altitudes below \SI{280}{\kilo\meter} using a drag-compensating propulsion system. As the altitude is reduced below \SI{400}{\kilo\meter} the propulsive and power requirements begin to increase and the mass correspondingly grows rapidly. When the potential of drag-reducing materials is incorporated into the design and propulsive and EPS performance is increased (Case VHR-2), operation at lower altitude can be achieved before the spacecraft mass begins to increase rapidly. This mass increase is shown in \cref{F:VHR_Subsystems}, as expected, to be due to the propulsion system. At higher altitude the mass of these two cases (VHR-1 and VHR-2) almost converge as a result of the diminishing requirement for drag compensation.

Designs with ABEP and lower performance input parameters (Case VHR-3) are shown to generally have a greater mass than an equivalent spacecraft with EP above approximately \SI{350}{\kilo\meter}. However, designs using these parameters are not found to converge for altitudes below \SI{300}{\kilo\meter} as a result of the assumed propulsive efficiency, power requirement, and material performance available at the current state-of-the-art. The power requirement of these systems is also high, indicated by the large solar array area required. However, when operating below \SI{500}{\kilo\meter}, these systems are still expected to be able to achieve a mass of at least half that of the most modern existing systems operating at higher altitude with a similar resolution (e.g. WorldView-4). When improved propulsive, intake, and aerodynamic material performance is incorporated (Case VHR-4), the total system mass reduces considerably and feasible designs are indicated at altitudes down to \SI{180}{\kilo\meter}. In \cref{F:VHR_Subsystems} the significant increase in power requirement of the propulsion system can be seen. Below approximately \SI{300}{\kilo\meter}, these designs are shown to have a mass lower than that of conventional EP systems despite lower thruster efficiency, demonstrating the significant potential of ABEP systems for sustained low-altitude operations in the future.

The altitudes identified for the designs of lowest mass with ABEP are somewhat higher than reflected in previous studies, which in some cases have suggested altitudes significantly lower than \SI{200}{\kilo\meter}. This is largely due to the geometric and aerodynamic models adopted in this work. Notably, both the core spacecraft body dimensions and the area of deployable solar arrays are included in the calculation of drag. Furthermore, the use of a GSI model that accounts for the interactions of surfaces that are nominally aligned with the oncoming flow also increases the expected drag. The result is such that the designs identified in this work, particularly at lower altitude, are slightly less optimistic.

Many of the previous and existing systems in orbit have been operated for a number of years beyond their initial planned lifetime (e.g. WorldView-1/2). However, as the orbital altitude is lowered, the possibility for extending operations also reduces for systems using EP for drag compensation as additional propellant is required to further delay orbital decay and re-entry. In comparison, for a longer initial lifetime requirement or extension of operations once in orbit, ABEP does not require additional propellant, though this does assume that the challenges of thruster development for long-term operations with atmospheric propellants have been addressed.

\cref{T:VHRout} also outlines some further top-level output parameters for the different cases and identified lowest-mass designs. It can be noted that the spacecraft body diameter for each case is significantly larger than the telescope aperture diameter required to achieve the target $\mathrm{GRD}$. This is due to the assumed use of existing telescope designs as a baseline for the system modelling process. The spacecraft body diameter is therefore related to the telescope length. However, a telescope for a slender spacecraft may be able to be designed specifically to minimise the cross-sectional area (with respect to the flow) through folding of the optical path, allowing configuration of the optical assembly along the length of the satellite body. This would first result in a reduction in drag with subsequent effect on the propulsive and power requirements and an associated reduction in total mass. This suggests that there may be scope to further develop and optimise such platforms for operation in VLEO with benefits to the overall system design.

The presented results also demonstrate that the cost of high performance optical VHR platforms generally reduces with orbital altitude, principally as a result of the reducing spacecraft size and mass. The cost of spacecraft operating in VLEO may therefore be significantly lower than comparable existing spacecraft that operate at higher altitude that have reported costs in the range of \SI{305}[\$]{\mega\nothing} to \SI{470}[\$]{\mega\nothing}. However, this comparison to existing systems is only indicative and it should be noted that the CERs utilised should be used primarily for relative rather than absolute analysis. Further cost savings associated with the use of lower altitude orbits not captured by this current analysis may however also be realised, for example due to the reduction in launch mass.

It should also be noted that systems operating at higher orbital altitude are likely to have larger access areas, greater overall coverage, and shorter revisit periods. Spacecraft operating at lower altitude, whilst individually smaller and cheaper, may therefore need to be operated in constellations to achieve the same total imaging output and performance. A further trade-off between the constellation design, performance, and total system costs therefore exists and requires exploration to understand the associated revenue and costs. Additional considerations in this trade-off may include aspects such as launch manifesting, constellation deployment, recurring and non-recurring engineering costs, and learning-curves.

\subsection{Synthetic Aperture Radar (SAR)}
Interest in synthetic aperture radar (SAR) systems continues to grow due to their ability to provide novel EO data products such as digital elevation mapping (DEM) and interferometric products that can provide detail information on geographical processes (e.g. terrain changes). SAR is also much less dependent on lighting conditions and cloud-cover and can therefore be used to provide imagery during night and poor weather. SAR is currently used predominantly in the defence and maritime sectors, but if made available at a lower cost would be attractive for a range of commercial applications such as construction and infrastructure monitoring, forestry surveying, agriculture, and disaster response \cite{Euroconsult2019}. Operation of a SAR satellite in VLEO may be advantageous due to the desired orientation of the SAR antenna. These payloads are typically aligned parallel to the direction of travel and therefore close to the direction of the oncoming flow, allowing a slender spacecraft geometry with a lower drag profile.

The principal mission and system parameters are provided in \cref{T:SARin}. A SAR satellite providing resolution comparative to current emerging market competitors (ICEYE and Capella Space) is envisaged. The input SAR antenna area density corresponds to the capability of modern SmallSat SAR concepts \cite{Foreman2016}.

\begin{table}
	\caption{SAR mission and system input parameters.} \label{T:SARin}
	\centering
	\begin{tabular}{l r}
	\toprule
	\bf{Parameter} & \bf{Value} \\
	\midrule
	Design lifetime [\si{years}] & 5 \\
	Azimuth Resolution (strip-map) [\si{\meter}] & 1.5 \\
	Range Resolution [\si{\meter}] & 0.25 \\
	Wavelength [\si{\milli\meter}] & 31.1 (X-band) \\
	Duty Cycle [-] & 0.25 \\
	NESZ [dB] & -20 \\
	SAR Antenna Area density [\si{\kilo\gram\per\meter\squared}] & 10\\
	Orbit & SSO \\
	LTAN & 15:00h \\
	\bottomrule
	\end{tabular}
	\end{table}

\begin{figure*}
	\centering
	\includegraphics[width=180mm]{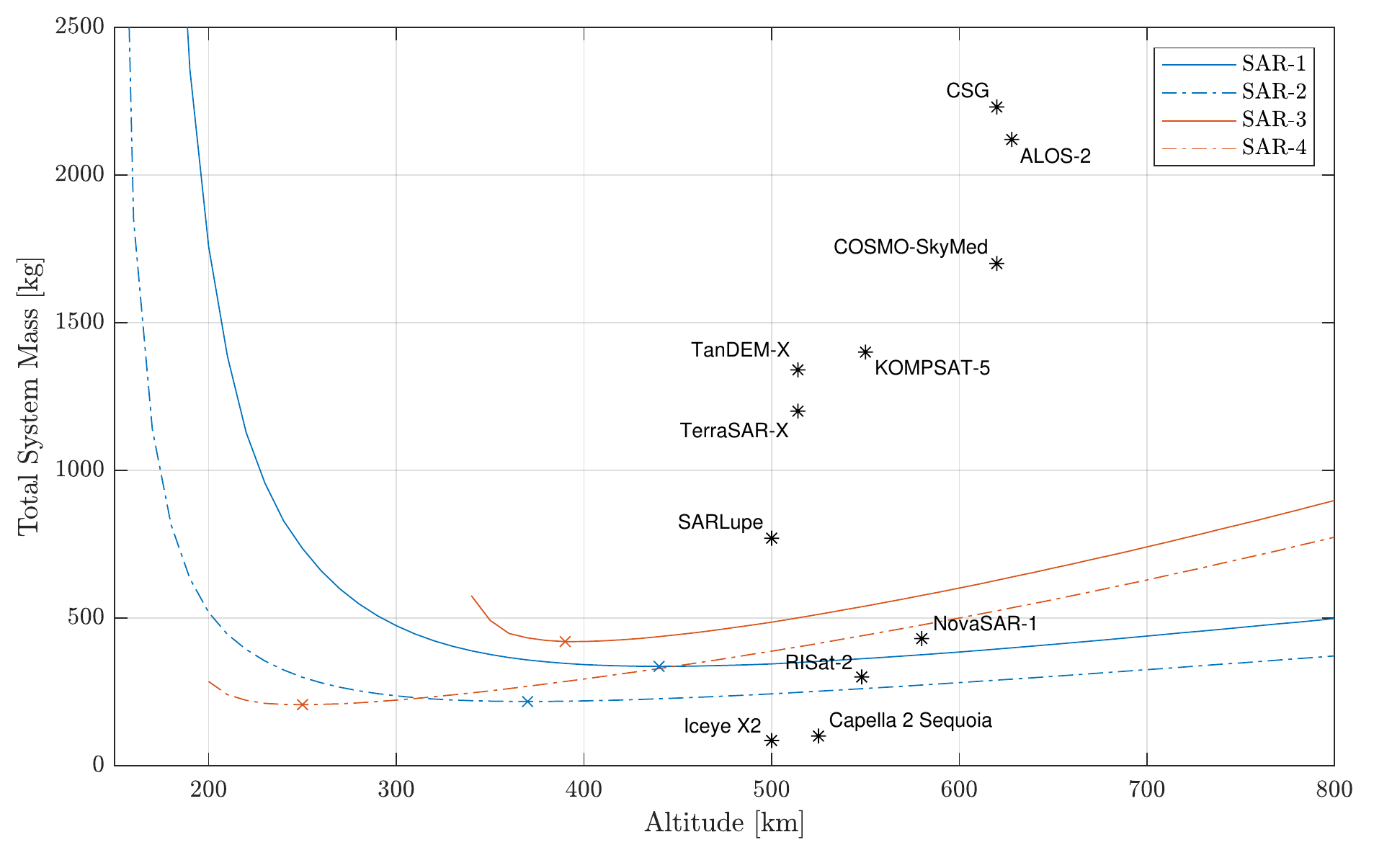}
	\caption{Variation of system mass with altitude for different SAR satellite designs. The point of minimum mass for each case is marked with by an 'x'. Previous SAR satellite designs are shown for comparison.}
	\label{F:SAR_Altitude}
	\end{figure*}

\begin{table*}
	\caption{SAR satellite design input and output parameters.} \label{T:SARout}
	\centering
	\begin{tabular}{l r r r r}
	\toprule
	\multirow{2}{*}{\bf{Parameter}} & \multicolumn{4}{c}{\bf{Case}} \\
	& SAR-1 & SAR-2 & SAR-3 & SAR-4 \\
	\midrule
	& \multicolumn{4}{c}{\bf{Inputs}} \\
	\midrule
	Propulsion Type 										& EP 	& EP 	& ABEP 	& ABEP 	\\
	Thrust Efficiency, $\eta_t$ [-]							& 0.60 	& 0.80 	& 0.30 	& 0.60 	\\
	Specific Impulse, $I_{sp}$ [$\si{\second}$]				& 2500	& 4000	& -		& -		\\
	Intake Efficiency, $\eta_c$ [-] 						& - 	& - 	& 0.40 	& 0.94 	\\
	Acc. Coefficient, $\alpha$ 	[-]							& 1.00 	& 0.50 	& 1.00 	& 0.50 	\\
	Solar Array Efficiency [-]								& 0.30	& 0.40	& 0.30	& 0.40	\\
	Solar Array Specific Power [\si{\watt\per\kilo\gram}]	& 60	& 100	& 60	& 100	\\
	Battery Energy Density [\si{\watt\hour\per\kilo\gram}]	& 125	& 200	& 125	& 200	\\
	\midrule
	& \multicolumn{4}{c}{\bf{Outputs}} \\
	\midrule
	Altitude [\si{\kilo\meter}] 							& 440 	& 370 	& 390 	& 250 	\\
	Mass [\si{\kilo\gram}] 									& 336 	& 217 	& 420 	& 206 	\\
	SAR Antenna Area [\si{\meter\squared}]					& 4.09 	& 3.44	& 3.62	& 2.32  \\
	Solar Array Area [\si{\meter\squared}]					& 17.4 	& 11.9 	& 21.1 	& 10.9	\\
	Unit Cost [\$M] 										& 345 	& 314 	& 334 	& 269 	\\
	\bottomrule
	\end{tabular}
	\end{table*}

For the additional parameters described in \cref{T:SARout}, output designs for SAR spacecraft for varying altitude have been generated and are shown in \cref{F:SAR_Altitude}. Compared to an optical payload, the mass of SAR instruments are less sensitive to increases in altitude and scale approximately linearly, from \SI{172}{\kilo\gram} at \SI{300}{\kilo\meter} to \SI{344}{\kilo\gram} at \SI{600}{\kilo\meter} for the given input requirements. However, the required power for the SAR instrument also increases considerably with range and therefore contributes to the increase in system mass at higher altitude despite the reduction in propulsive requirement. This is particularly notable for Case SAR-1 and SAR-3 that have lower input EPS performance parameters.

In general, the total mass presented for the designed systems is lower than the majority of previous high performing SAR satellites and similar to that of the modern small SAR satellites such as ICEYE and Capella Space \cite{Euroconsult2019}. This is principally due to the low antenna area density value provided as input that corresponds to fewer imaging modes and therefore results in lower mass solutions. This is consistent with the trend identified by \citet{Paek2020} of moving away from conventional design principles of large monolithic spacecraft with dedicated service to specific customer segments towards constellations of smaller satellites aimed at wider commercial markets.

At lower altitude, the total system mass is shown to increase significantly as the propulsive requirements associated with maintaining the mission orbit for the desired lifetime increase. For the presented inputs, the ABEP system with drag-reducing materials (Case SAR-4) provides the lowest mass solution at an altitude of \SI{250}{\kilo\meter}, though the sensitivity of the different solutions to altitude is much smaller than the previous case study for the VHR optical payload. In this case the ABEP solutions are shown to be unable to converge for altitudes below approximately \SI{425}{\kilo\meter} (Case SAR-3) and \SI{275}{\kilo\meter} (Case SAR-4).

Some previous SAR systems have utilised hexagonal body configurations (e.g. TerraSAR-X and TanDEM-X) that exploit the shape and required orientation of the antenna to generally reduce the experienced drag whilst conforming to traditional design principles. However, more recent designs have considered more radical slender configurations (e.g. ICEYE X2 and NovaSAR) or alternative deployable antenna designs (e.g. Capella Space Sequoia) that may further reduce the projected area of the spacecraft with respect to the flow. The approximately cylindrical geometry adopted for this study may therefore be a disadvantage and alternative configurations could be considered that can further benefit the overall system design, particularly if novel aerodynamic materials become available.

Whilst the reported cost estimates suggest a general reduction in cost of operating at lower altitude, they remain significantly higher than that of the emerging competitors, for example ICEYE (\$7M), Capella Space (\$8M), and even that of TerraSAR-X/TanDEM-X (\$117M). They are however commensurate with the earlier monolithic SAR spacecraft, e.g. Radarsat-2 (\$433M) and COSMO-SkyMed (\$364M). This is principally attributed to the use of the NASA instrument cost model that significantly overestimates the SAR payload cost in comparison to modern commercial instruments and accounts for \SIrange{50}{70}{\%} of the total spacecraft cost. The cost of \emph{Telemetry, Tracking, Control, and Data Handling} from the SSCM accounts for a further \SIrange{15}{30}{\%} of the total. These results suggest that the available CERs do not capture the recent advancements in this market segment to reflect the modern design and development principles of commercial small satellites, particularly with SAR payloads. This further suggests that in this context the use of cost models and CERs should be used carefully and for relative rather than absolute indication of expected costs. 

\section{Conclusions}
New engineering system models for emerging VLEO technologies have been developed and applied to the conceptual design of spacecraft for EO applications. Through integration with conventional system models for spacecraft design, the trends and trade-offs associated with the use of these technologies for sustained operations at reducing orbital altitude can be explored.

The use of both conventional EP and novel ABEP for drag compensation were shown to enable operations at significantly lower altitude than current commercial EO spacecraft. The benefit to payload design at lower altitude was shown to result in a significant reduction in the total system mass, though this was bounded at very low altitudes by the increasing atmospheric density and drag. The use of drag-reducing materials, enhanced electrical power systems, and higher thruster efficiency were shown to further increase the system performance and enable further reductions in orbital altitude, resulting in spacecraft of even lower system mass. 

The results of two case-studies, an optical VHR satellite and a SAR satellite, demonstrated the reduction in the size and mass of EO spacecraft that could be achieved by designing for operations in lower altitude orbits whilst maintaining or improving observational performance. It was also indicated, using available cost models, that substantial savings in system development and manufacture cost could be achieved for a notional optical VHR satellite. In contrast, cost savings of this magnitude were not observed for the SAR satellite, though this result was largely attributed to the use of CERs that do not reflect the modern development of commercial SAR instruments and small satellites.

The presented results demonstrate the significant potential of VLEO, particularly for the development of EO missions with increasing performance requirements. A range of further benefits of operating at lower orbital altitude exist that are yet to be accounted for in the system modelling framework, for example increased launch vehicle capability, reduced radiation exposure, and lower collision risk. Further mission and system design explorations are also required to investigate the trade-offs associated with the design and performance of constellations of EO spacecraft operating in VLEO. The use of elliptical orbits with a low perigee altitude may also be considered in these explorations to provide improved regional coverage whilst also benefiting from reduced propulsive requirements.

The development of this systems modelling framework has demonstrated the need for a more holistic approach to conceptual spacecraft design for VLEO that can capture the complex interactions involved, particularly when the effect of novel technologies are are being considered. As they are still in development, significant uncertainty remains in the performance estimation and route to implementation of these enabling technologies for VLEO. The approach to system modelling presented in this paper lays out the means to explore the impact of these technologies on the spacecraft design, set requirements for their development, and inform their ongoing testing and implementation towards enabling sustainable VLEO spacecraft in the future.

\section*{Acknowledgements}
This project has received funding from the European Union's Horizon 2020 research and innovation programme under grant agreement No 737183. This publication reflects only the view of the authors. The European Commission is not responsible for any use that may be made of the information it contains.

\bibliography{VLEO_Systems_Modelling}

\begin{thebibliography}{97}
\expandafter\ifx\csname natexlab\endcsname\relax\def\natexlab#1{#1}\fi
\providecommand{\url}[1]{\texttt{#1}}
\providecommand{\href}[2]{#2}
\providecommand{\path}[1]{#1}
\providecommand{\DOIprefix}{doi:}
\providecommand{\ArXivprefix}{arXiv:}
\providecommand{\URLprefix}{URL: }
\providecommand{\Pubmedprefix}{pmid:}
\providecommand{\doi}[1]{\href{http://dx.doi.org/#1}{\path{#1}}}
\providecommand{\Pubmed}[1]{\href{pmid:#1}{\path{#1}}}
\providecommand{\bibinfo}[2]{#2}
\ifx\xfnm\relax \def\xfnm[#1]{\unskip,\space#1}\fi
\bibitem[{Kansakar and Hossain(2016)}]{Kansakar2016}
\bibinfo{author}{P.~Kansakar}, \bibinfo{author}{F.~Hossain},
\newblock \bibinfo{title}{{A review of applications of satellite earth
  observation data for global societal benefit and stewardship of planet
  earth}},
\newblock \bibinfo{journal}{Space Policy} \bibinfo{volume}{36}
  (\bibinfo{year}{2016}) \bibinfo{pages}{46--54}. \URLprefix
  \url{http://dx.doi.org/10.1016/j.spacepol.2016.05.005}.
  \DOIprefix\doi{10.1016/j.spacepol.2016.05.005}.
\bibitem[{Euroconsult(2019)}]{Euroconsult2019}
\bibinfo{author}{Euroconsult}, \bibinfo{title}{{Satellite-Based Earth
  Observation Market Prospects to 2028}}, \bibinfo{type}{Technical Report},
  \bibinfo{address}{Paris, France}, \bibinfo{year}{2019}.
\bibitem[{Wertz et~al.(2012)Wertz, Sarzi-Amad{\'{e}}, Shao, Taylor, and {Van
  Allen}}]{Wertz2012a}
\bibinfo{author}{J.~R. Wertz}, \bibinfo{author}{N.~Sarzi-Amad{\'{e}}},
  \bibinfo{author}{A.~E. Shao}, \bibinfo{author}{C.~Taylor},
  \bibinfo{author}{R.~{Van Allen}},
\newblock \bibinfo{title}{{Moderately Elliptical Very Low Orbits (MEVLOs) as a
  Long-Term Solution to Orbital Debris}},
\newblock \bibinfo{journal}{26th Annual AIAA/USU Conference on Small
  Satellites, Logan, UT}  (\bibinfo{year}{2012}).
\bibitem[{Roberts et~al.(2017)Roberts, Crisp, Edmondson, Haigh, Lyons, Oiko,
  Macario-Rojas, Smith, Becedas, Gonz{\'{a}}lez, V{\'{a}}zquez, Bra{\~{n}}a,
  Antonini, Bay, Ghizoni, Jungnell, Morsb{\o}l, Binder, Boxberger, Herdrich,
  Romano, Fasoulas, Garcia-Almi{\~{n}}ana, Rodriguez-Donaire, Kataria,
  Davidson, Outlaw, Belkouchi, Conte, Perez, Villain, Hei{\ss}erer, and
  Schwalber}]{Roberts2017}
\bibinfo{author}{P.~C. Roberts}, \bibinfo{author}{N.~H. Crisp},
  \bibinfo{author}{S.~Edmondson}, \bibinfo{author}{S.~J. Haigh},
  \bibinfo{author}{R.~E. Lyons}, \bibinfo{author}{V.~T. Oiko},
  \bibinfo{author}{A.~Macario-Rojas}, \bibinfo{author}{K.~L. Smith},
  \bibinfo{author}{J.~Becedas}, \bibinfo{author}{G.~Gonz{\'{a}}lez},
  \bibinfo{author}{I.~V{\'{a}}zquez}, \bibinfo{author}{{\'{A}}.~Bra{\~{n}}a},
  \bibinfo{author}{K.~Antonini}, \bibinfo{author}{K.~Bay},
  \bibinfo{author}{L.~Ghizoni}, \bibinfo{author}{V.~Jungnell},
  \bibinfo{author}{J.~Morsb{\o}l}, \bibinfo{author}{T.~Binder},
  \bibinfo{author}{A.~Boxberger}, \bibinfo{author}{G.~H. Herdrich},
  \bibinfo{author}{F.~Romano}, \bibinfo{author}{S.~Fasoulas},
  \bibinfo{author}{D.~Garcia-Almi{\~{n}}ana},
  \bibinfo{author}{S.~Rodriguez-Donaire}, \bibinfo{author}{D.~Kataria},
  \bibinfo{author}{M.~Davidson}, \bibinfo{author}{R.~Outlaw},
  \bibinfo{author}{B.~Belkouchi}, \bibinfo{author}{A.~Conte},
  \bibinfo{author}{J.~S. Perez}, \bibinfo{author}{R.~Villain},
  \bibinfo{author}{B.~Hei{\ss}erer}, \bibinfo{author}{A.~Schwalber},
\newblock \bibinfo{title}{{DISCOVERER – Radical Redesign of Earth Observation
  Satellites for Sustained Operation at Significantly Lower Altitudes}},
\newblock in: \bibinfo{booktitle}{68th International Astronautical Congress},
  \bibinfo{number}{September}, \bibinfo{publisher}{International Astronautical
  Federation (IAF)}, \bibinfo{address}{Adelaide, Australia},
  \bibinfo{year}{2017}, pp. \bibinfo{pages}{1--9}.
\bibitem[{Roberts et~al.(2019)Roberts, Crisp, Romano, Herdrich, Oiko,
  Edmondson, Haigh, Huyton, Livadiotti, Lyons, Smith, Sinpetru, Straker,
  Worrall, Becedas, Dom{\'{i}}nguez, Gonz{\'{a}}lez, Ca{\~{n}}as, Hanessian,
  M{\o}lgaard, Nielsen, Bisgaard, Boxberger, Chan, Fasoulas, Traub,
  Garcia-Almi{\~{n}}ana, Rodriguez-Donaire, Sureda, Kataria, Outlaw, Belkouchi,
  Conte, Perez, Villain, Hei{\ss}erer, and Schwalber}]{Roberts2019}
\bibinfo{author}{P.~C. Roberts}, \bibinfo{author}{N.~H. Crisp},
  \bibinfo{author}{F.~Romano}, \bibinfo{author}{G.~H. Herdrich},
  \bibinfo{author}{V.~T. Oiko}, \bibinfo{author}{S.~Edmondson},
  \bibinfo{author}{S.~J. Haigh}, \bibinfo{author}{C.~Huyton},
  \bibinfo{author}{S.~Livadiotti}, \bibinfo{author}{R.~E. Lyons},
  \bibinfo{author}{K.~L. Smith}, \bibinfo{author}{L.~A. Sinpetru},
  \bibinfo{author}{A.~Straker}, \bibinfo{author}{S.~D. Worrall},
  \bibinfo{author}{J.~Becedas}, \bibinfo{author}{R.~M. Dom{\'{i}}nguez},
  \bibinfo{author}{D.~Gonz{\'{a}}lez}, \bibinfo{author}{V.~Ca{\~{n}}as},
  \bibinfo{author}{V.~Hanessian}, \bibinfo{author}{A.~M{\o}lgaard},
  \bibinfo{author}{J.~Nielsen}, \bibinfo{author}{M.~Bisgaard},
  \bibinfo{author}{A.~Boxberger}, \bibinfo{author}{Y.-A. Chan},
  \bibinfo{author}{S.~Fasoulas}, \bibinfo{author}{C.~Traub},
  \bibinfo{author}{D.~Garcia-Almi{\~{n}}ana},
  \bibinfo{author}{S.~Rodriguez-Donaire}, \bibinfo{author}{M.~Sureda},
  \bibinfo{author}{D.~Kataria}, \bibinfo{author}{R.~Outlaw},
  \bibinfo{author}{B.~Belkouchi}, \bibinfo{author}{A.~Conte},
  \bibinfo{author}{J.~S. Perez}, \bibinfo{author}{R.~Villain},
  \bibinfo{author}{B.~Hei{\ss}erer}, \bibinfo{author}{A.~Schwalber},
\newblock \bibinfo{title}{{DISCOVERER – Making Commercial Satellite
  Operations in Very Low Earth Orbit a Reality}},
\newblock in: \bibinfo{booktitle}{70th International Astronautical Congress},
  \bibinfo{publisher}{International Astronautical Federation (IAF)},
  \bibinfo{address}{Washington, DC}, \bibinfo{year}{2019}, pp.
  \bibinfo{pages}{1--9}.
\bibitem[{Roberts et~al.(2020)Roberts, Crisp, Edmondson, Haigh, Holmes,
  Livadiotti, Macario-Rojas, Oiko, Smith, Sinpetru, Becedas, Dom{\'{i}}nguez,
  Sulliotti-Linner, Christensen, Jensen, Nielsen, Bisgaard, Chan, Herdrich,
  Romano, Fasoulas, Traub, Garcia-Almi{\~{n}}ana, Garcia-Berenguer,
  Rodriguez-Donaire, Sureda, Kataria, Belkouchi, Conte, Seminari, Villain, and
  Schwalber}]{Roberts2020}
\bibinfo{author}{P.~C. Roberts}, \bibinfo{author}{N.~H. Crisp},
  \bibinfo{author}{S.~Edmondson}, \bibinfo{author}{S.~J. Haigh},
  \bibinfo{author}{B.~E. Holmes}, \bibinfo{author}{S.~Livadiotti},
  \bibinfo{author}{A.~Macario-Rojas}, \bibinfo{author}{V.~T. Oiko},
  \bibinfo{author}{K.~L. Smith}, \bibinfo{author}{L.~A. Sinpetru},
  \bibinfo{author}{J.~Becedas}, \bibinfo{author}{R.~M. Dom{\'{i}}nguez},
  \bibinfo{author}{V.~Sulliotti-Linner}, \bibinfo{author}{S.~Christensen},
  \bibinfo{author}{T.~K. Jensen}, \bibinfo{author}{J.~Nielsen},
  \bibinfo{author}{M.~Bisgaard}, \bibinfo{author}{Y.-A. Chan},
  \bibinfo{author}{G.~H. Herdrich}, \bibinfo{author}{F.~Romano},
  \bibinfo{author}{S.~Fasoulas}, \bibinfo{author}{C.~Traub},
  \bibinfo{author}{D.~Garcia-Almi{\~{n}}ana},
  \bibinfo{author}{M.~Garcia-Berenguer},
  \bibinfo{author}{S.~Rodriguez-Donaire}, \bibinfo{author}{M.~Sureda},
  \bibinfo{author}{D.~Kataria}, \bibinfo{author}{B.~Belkouchi},
  \bibinfo{author}{A.~Conte}, \bibinfo{author}{S.~Seminari},
  \bibinfo{author}{R.~Villain}, \bibinfo{author}{A.~Schwalber},
\newblock \bibinfo{title}{{DISCOVERER: Developing Technologies to Enable
  Commercial Satellite Operations in Very Low Earth Orbit}},
\newblock in: \bibinfo{booktitle}{71st International Astronautical Congress –
  The CyberSpace Edition}, \bibinfo{publisher}{International Astronautical
  Federation (IAF)}, \bibinfo{year}{2020}.
\bibitem[{Crisp et~al.(2020)Crisp, Roberts, Livadiotti, Oiko, Edmondson, Haigh,
  Huyton, Sinpetru, Smith, Worrall, Becedas, Dom{\'{i}}nguez, Gonz{\'{a}}lez,
  Hanessian, M{\o}lgaard, Nielsen, Bisgaard, Chan, Fasoulas, Herdrich, Romano,
  Traub, Garc{\'{i}}a-Almi{\~{n}}ana, Rodr{\'{i}}guez-Donaire, Sureda, Kataria,
  Outlaw, Belkouchi, Conte, Perez, Villain, Hei{\ss}erer, and
  Schwalber}]{Crisp2020}
\bibinfo{author}{N.~Crisp}, \bibinfo{author}{P.~Roberts},
  \bibinfo{author}{S.~Livadiotti}, \bibinfo{author}{V.~Oiko},
  \bibinfo{author}{S.~Edmondson}, \bibinfo{author}{S.~Haigh},
  \bibinfo{author}{C.~Huyton}, \bibinfo{author}{L.~Sinpetru},
  \bibinfo{author}{K.~Smith}, \bibinfo{author}{S.~Worrall},
  \bibinfo{author}{J.~Becedas}, \bibinfo{author}{R.~Dom{\'{i}}nguez},
  \bibinfo{author}{D.~Gonz{\'{a}}lez}, \bibinfo{author}{V.~Hanessian},
  \bibinfo{author}{A.~M{\o}lgaard}, \bibinfo{author}{J.~Nielsen},
  \bibinfo{author}{M.~Bisgaard}, \bibinfo{author}{Y.-A. Chan},
  \bibinfo{author}{S.~Fasoulas}, \bibinfo{author}{G.~Herdrich},
  \bibinfo{author}{F.~Romano}, \bibinfo{author}{C.~Traub},
  \bibinfo{author}{D.~Garc{\'{i}}a-Almi{\~{n}}ana},
  \bibinfo{author}{S.~Rodr{\'{i}}guez-Donaire}, \bibinfo{author}{M.~Sureda},
  \bibinfo{author}{D.~Kataria}, \bibinfo{author}{R.~Outlaw},
  \bibinfo{author}{B.~Belkouchi}, \bibinfo{author}{A.~Conte},
  \bibinfo{author}{J.~Perez}, \bibinfo{author}{R.~Villain},
  \bibinfo{author}{B.~Hei{\ss}erer}, \bibinfo{author}{A.~Schwalber},
\newblock \bibinfo{title}{{The benefits of very low earth orbit for earth
  observation missions}},
\newblock \bibinfo{journal}{Progress in Aerospace Sciences}
  \bibinfo{volume}{117} (\bibinfo{year}{2020}) \bibinfo{pages}{100619}.
  \URLprefix
  \url{https://linkinghub.elsevier.com/retrieve/pii/S0376042120300312}.
  \DOIprefix\doi{10.1016/j.paerosci.2020.100619}.
\bibitem[{Richelson(1984)}]{Richelson1984}
\bibinfo{author}{J.~Richelson},
\newblock \bibinfo{title}{{The keyhole satellite program}},
\newblock \bibinfo{journal}{Journal of Strategic Studies} \bibinfo{volume}{7}
  (\bibinfo{year}{1984}) \bibinfo{pages}{121--153}. \URLprefix
  \url{http://www.tandfonline.com/doi/abs/10.1080/01402398408437182}.
  \DOIprefix\doi{10.1080/01402398408437182}.
\bibitem[{Steiger et~al.(2014)Steiger, Romanazzo, Emanuelli, Floberghagen, and
  Fehringer}]{Steiger2014}
\bibinfo{author}{C.~Steiger}, \bibinfo{author}{M.~Romanazzo},
  \bibinfo{author}{P.~P. Emanuelli}, \bibinfo{author}{R.~Floberghagen},
  \bibinfo{author}{M.~Fehringer},
\newblock \bibinfo{title}{{The Deorbiting of ESA's Gravity Mission GOCE -
  Spacecraft Operations in Extreme Drag Conditions}},
\newblock in: \bibinfo{booktitle}{SpaceOps 2014 Conference},
  \bibinfo{number}{May}, \bibinfo{publisher}{American Institute of Aeronautics
  and Astronautics}, \bibinfo{address}{Reston, Virginia}, \bibinfo{year}{2014},
  pp. \bibinfo{pages}{1--13}. \URLprefix
  \url{http://arc.aiaa.org/doi/10.2514/6.2014-1934}.
  \DOIprefix\doi{10.2514/6.2014-1934}.
\bibitem[{Picone et~al.(2002)Picone, Hedin, Drob, and Aikin}]{Picone2002}
\bibinfo{author}{J.~Picone}, \bibinfo{author}{A.~Hedin}, \bibinfo{author}{D.~P.
  Drob}, \bibinfo{author}{A.~Aikin},
\newblock \bibinfo{title}{{NRLMSISE-00 Empirical Model of the Atmosphere:
  Statistical Comparisons and Scientific Issues}},
\newblock \bibinfo{journal}{Journal of Geophysical Research}
  \bibinfo{volume}{107} (\bibinfo{year}{2002}).
  \DOIprefix\doi{10.1029/2002JA009430}.
\bibitem[{Sentman(1961)}]{Sentman1961}
\bibinfo{author}{L.~H. Sentman}, \bibinfo{title}{{Free molecule flow theory and
  its application to the determination of aerodynamic forces}},
  \bibinfo{type}{Technical Report}, Lockheed Missiles {\&} Space Company,
  \bibinfo{address}{Sunnyvale, CA}, \bibinfo{year}{1961}.
\bibitem[{Moe et~al.(1998)Moe, Moe, and Wallace}]{Moe1998}
\bibinfo{author}{K.~Moe}, \bibinfo{author}{M.~M. Moe}, \bibinfo{author}{S.~D.
  Wallace},
\newblock \bibinfo{title}{{Improved Satellite Drag Coefficient Calculations
  from Orbital Measurements of Energy Accommodation}},
\newblock \bibinfo{journal}{Journal of Spacecraft and Rockets}
  \bibinfo{volume}{35} (\bibinfo{year}{1998}) \bibinfo{pages}{266--272}.
  \URLprefix \url{http://arc.aiaa.org/doi/10.2514/2.3350}.
  \DOIprefix\doi{10.2514/2.3350}.
\bibitem[{Mostaza-Prieto et~al.(2014)Mostaza-Prieto, Graziano, and
  Roberts}]{MostazaPrieto2014}
\bibinfo{author}{D.~Mostaza-Prieto}, \bibinfo{author}{B.~P. Graziano},
  \bibinfo{author}{P.~C. Roberts},
\newblock \bibinfo{title}{{Spacecraft drag modelling}},
\newblock \bibinfo{journal}{Progress in Aerospace Sciences}
  \bibinfo{volume}{64} (\bibinfo{year}{2014}) \bibinfo{pages}{56--65}.
  \URLprefix \url{http://dx.doi.org/10.1016/j.paerosci.2013.09.001}.
  \DOIprefix\doi{10.1016/j.paerosci.2013.09.001}.
\bibitem[{Livadiotti et~al.(2020)Livadiotti, Crisp, Roberts, Worrall, Oiko,
  Edmondson, Haigh, Huyton, Smith, Sinpetru, Holmes, Becedas, Dom{\'{i}}nguez,
  Ca{\~{n}}as, Christensen, M{\o}lgaard, Nielsen, Bisgaard, Chan, Herdrich,
  Romano, Fasoulas, Traub, Garcia-Almi{\~{n}}ana, Rodriguez-Donaire, Sureda,
  Kataria, Belkouchi, Conte, Perez, Villain, and Outlaw}]{Livadiotti2020}
\bibinfo{author}{S.~Livadiotti}, \bibinfo{author}{N.~H. Crisp},
  \bibinfo{author}{P.~C. Roberts}, \bibinfo{author}{S.~D. Worrall},
  \bibinfo{author}{V.~T. Oiko}, \bibinfo{author}{S.~Edmondson},
  \bibinfo{author}{S.~J. Haigh}, \bibinfo{author}{C.~Huyton},
  \bibinfo{author}{K.~L. Smith}, \bibinfo{author}{L.~A. Sinpetru},
  \bibinfo{author}{B.~E. Holmes}, \bibinfo{author}{J.~Becedas},
  \bibinfo{author}{R.~M. Dom{\'{i}}nguez}, \bibinfo{author}{V.~Ca{\~{n}}as},
  \bibinfo{author}{S.~Christensen}, \bibinfo{author}{A.~M{\o}lgaard},
  \bibinfo{author}{J.~Nielsen}, \bibinfo{author}{M.~Bisgaard},
  \bibinfo{author}{Y.-A. Chan}, \bibinfo{author}{G.~H. Herdrich},
  \bibinfo{author}{F.~Romano}, \bibinfo{author}{S.~Fasoulas},
  \bibinfo{author}{C.~Traub}, \bibinfo{author}{D.~Garcia-Almi{\~{n}}ana},
  \bibinfo{author}{S.~Rodriguez-Donaire}, \bibinfo{author}{M.~Sureda},
  \bibinfo{author}{D.~Kataria}, \bibinfo{author}{B.~Belkouchi},
  \bibinfo{author}{A.~Conte}, \bibinfo{author}{J.~S. Perez},
  \bibinfo{author}{R.~Villain}, \bibinfo{author}{R.~Outlaw},
\newblock \bibinfo{title}{{A review of gas-surface interaction models for
  orbital aerodynamics applications}},
\newblock \bibinfo{journal}{Progress in Aerospace Sciences}
  \bibinfo{volume}{119} (\bibinfo{year}{2020}) \bibinfo{pages}{100675}.
  \URLprefix
  \url{https://linkinghub.elsevier.com/retrieve/pii/S0376042120300877}.
  \DOIprefix\doi{10.1016/j.paerosci.2020.100675}.
\bibitem[{Banks et~al.(2004)Banks, Miller, and de~Groh}]{Banks2004}
\bibinfo{author}{B.~Banks}, \bibinfo{author}{S.~Miller},
  \bibinfo{author}{K.~de~Groh},
\newblock \bibinfo{title}{{Low Earth Orbital Atomic Oxygen Interactions with
  Materials}},
\newblock in: \bibinfo{booktitle}{2nd International Energy Conversion
  Engineering Conference}, \bibinfo{publisher}{American Institute of
  Aeronautics and Astronautics}, \bibinfo{address}{Providence, RI},
  \bibinfo{year}{2004}. \URLprefix
  \url{http://arc.aiaa.org/doi/10.2514/6.2004-5638}.
  \DOIprefix\doi{10.2514/6.2004-5638}.
\bibitem[{Pardini et~al.(2010)Pardini, Anselmo, Moe, and Moe}]{Pardini2010}
\bibinfo{author}{C.~Pardini}, \bibinfo{author}{L.~Anselmo},
  \bibinfo{author}{K.~Moe}, \bibinfo{author}{M.~M. Moe},
\newblock \bibinfo{title}{{Drag and energy accommodation coefficients during
  sunspot maximum}},
\newblock \bibinfo{journal}{Advances in Space Research} \bibinfo{volume}{45}
  (\bibinfo{year}{2010}) \bibinfo{pages}{638--650}. \URLprefix
  \url{http://dx.doi.org/10.1016/j.asr.2009.08.034}.
  \DOIprefix\doi{10.1016/j.asr.2009.08.034}.
\bibitem[{Macario-Rojas et~al.(2018)Macario-Rojas, Smith, Crisp, and
  Roberts}]{Macario-Rojas2018}
\bibinfo{author}{A.~Macario-Rojas}, \bibinfo{author}{K.~L. Smith},
  \bibinfo{author}{N.~H. Crisp}, \bibinfo{author}{P.~C. Roberts},
\newblock \bibinfo{title}{{Atmospheric interaction with nanosatellites from
  observed orbital decay}},
\newblock \bibinfo{journal}{Advances in Space Research} \bibinfo{volume}{61}
  (\bibinfo{year}{2018}) \bibinfo{pages}{2972--2982}.
  \DOIprefix\doi{10.1016/j.asr.2018.02.022}.
\bibitem[{Emmert(2015)}]{Emmert2015}
\bibinfo{author}{J.~Emmert},
\newblock \bibinfo{title}{{Thermospheric mass density: A review}},
\newblock \bibinfo{journal}{Advances in Space Research} \bibinfo{volume}{56}
  (\bibinfo{year}{2015}) \bibinfo{pages}{773--824}. \URLprefix
  \url{http://dx.doi.org/10.1016/j.asr.2015.05.038}.
  \DOIprefix\doi{10.1016/j.asr.2015.05.038}.
\bibitem[{Singh and Walker(2015)}]{Singh2015}
\bibinfo{author}{L.~A. Singh}, \bibinfo{author}{M.~L. Walker},
\newblock \bibinfo{title}{{A review of research in low earth orbit propellant
  collection}},
\newblock \bibinfo{journal}{Progress in Aerospace Sciences}
  \bibinfo{volume}{75} (\bibinfo{year}{2015}) \bibinfo{pages}{15--25}.
  \URLprefix \url{http://dx.doi.org/10.1016/j.paerosci.2015.03.001
  https://linkinghub.elsevier.com/retrieve/pii/S0376042115000226}.
  \DOIprefix\doi{10.1016/j.paerosci.2015.03.001}.
\bibitem[{Sch{\"{o}}nherr et~al.(2015)Sch{\"{o}}nherr, Komurasaki, Romano,
  Massuti-Ballester, and Herdrich}]{Schonherr2015}
\bibinfo{author}{T.~Sch{\"{o}}nherr}, \bibinfo{author}{K.~Komurasaki},
  \bibinfo{author}{F.~Romano}, \bibinfo{author}{B.~Massuti-Ballester},
  \bibinfo{author}{G.~H. Herdrich},
\newblock \bibinfo{title}{{Analysis of atmosphere-breathing electric
  propulsion}},
\newblock \bibinfo{journal}{IEEE Transactions on Plasma Science}
  \bibinfo{volume}{43} (\bibinfo{year}{2015}) \bibinfo{pages}{287--294}.
  \DOIprefix\doi{10.1109/TPS.2014.2364053}.
\bibitem[{Romano et~al.(2020)Romano, Chan, Herdrich, Traub, Fasoulas, Roberts,
  Smith, Edmondson, Haigh, Crisp, Oiko, Worrall, Livadiotti, Huyton, Sinpetru,
  Straker, Becedas, Dom{\'{i}}nguez, Gonz{\'{a}}lez, Ca{\~{n}}as,
  Sulliotti-Linner, Hanessian, M{\o}lgaard, Nielsen, Bisgaard,
  Garcia-Almi{\~{n}}ana, Rodriguez-Donaire, Sureda, Kataria, Outlaw, Villain,
  Perez, Conte, Belkouchi, Schwalber, and Hei{\ss}erer}]{Romano2020}
\bibinfo{author}{F.~Romano}, \bibinfo{author}{Y.-A. Chan},
  \bibinfo{author}{G.~Herdrich}, \bibinfo{author}{C.~Traub},
  \bibinfo{author}{S.~Fasoulas}, \bibinfo{author}{P.~Roberts},
  \bibinfo{author}{K.~Smith}, \bibinfo{author}{S.~Edmondson},
  \bibinfo{author}{S.~Haigh}, \bibinfo{author}{N.~Crisp},
  \bibinfo{author}{V.~Oiko}, \bibinfo{author}{S.~Worrall},
  \bibinfo{author}{S.~Livadiotti}, \bibinfo{author}{C.~Huyton},
  \bibinfo{author}{L.~Sinpetru}, \bibinfo{author}{A.~Straker},
  \bibinfo{author}{J.~Becedas}, \bibinfo{author}{R.~Dom{\'{i}}nguez},
  \bibinfo{author}{D.~Gonz{\'{a}}lez}, \bibinfo{author}{V.~Ca{\~{n}}as},
  \bibinfo{author}{V.~Sulliotti-Linner}, \bibinfo{author}{V.~Hanessian},
  \bibinfo{author}{A.~M{\o}lgaard}, \bibinfo{author}{J.~Nielsen},
  \bibinfo{author}{M.~Bisgaard}, \bibinfo{author}{D.~Garcia-Almi{\~{n}}ana},
  \bibinfo{author}{S.~Rodriguez-Donaire}, \bibinfo{author}{M.~Sureda},
  \bibinfo{author}{D.~Kataria}, \bibinfo{author}{R.~Outlaw},
  \bibinfo{author}{R.~Villain}, \bibinfo{author}{J.~Perez},
  \bibinfo{author}{A.~Conte}, \bibinfo{author}{B.~Belkouchi},
  \bibinfo{author}{A.~Schwalber}, \bibinfo{author}{B.~Hei{\ss}erer},
\newblock \bibinfo{title}{{RF Helicon-based Inductive Plasma Thruster (IPT)
  Design for an Atmosphere-Breathing Electric Propulsion system (ABEP)}},
\newblock \bibinfo{journal}{Acta Astronautica} \bibinfo{volume}{176}
  (\bibinfo{year}{2020}) \bibinfo{pages}{476--483}. \URLprefix
  \url{https://linkinghub.elsevier.com/retrieve/pii/S0094576520304264}.
  \DOIprefix\doi{10.1016/j.actaastro.2020.07.008}.
\bibitem[{Guelman and Kogan(1999)}]{Guelman1999}
\bibinfo{author}{M.~Guelman}, \bibinfo{author}{A.~Kogan},
\newblock \bibinfo{title}{{Electric Propulsion for Remote Sensing from Low
  Orbits}},
\newblock \bibinfo{journal}{Journal of Guidance, Control, and Dynamics}
  \bibinfo{volume}{22} (\bibinfo{year}{1999}) \bibinfo{pages}{313--321}.
  \URLprefix \url{https://arc.aiaa.org/doi/10.2514/2.4380}.
  \DOIprefix\doi{10.2514/2.4380}.
\bibitem[{Co and Black(2014)}]{Co2014}
\bibinfo{author}{T.~C. Co}, \bibinfo{author}{J.~T. Black},
\newblock \bibinfo{title}{{Responsiveness in Low Orbits Using Electric
  Propulsion}},
\newblock \bibinfo{journal}{Journal of Spacecraft and Rockets}
  \bibinfo{volume}{51} (\bibinfo{year}{2014}) \bibinfo{pages}{938--945}.
  \DOIprefix\doi{10.2514/1.A32405}.
\bibitem[{Guelman and Shiryaev(2019)}]{Guelman2019}
\bibinfo{author}{M.~M. Guelman}, \bibinfo{author}{A.~Shiryaev},
\newblock \bibinfo{title}{{Closed-Loop Control of Earth Observation
  Satellites}},
\newblock \bibinfo{journal}{Journal of Spacecraft and Rockets}
  \bibinfo{volume}{56} (\bibinfo{year}{2019}) \bibinfo{pages}{82--90}.
  \URLprefix \url{https://arc.aiaa.org/doi/10.2514/1.A34134}.
  \DOIprefix\doi{10.2514/1.A34134}.
\bibitem[{Leppinen(2016)}]{Leppinen2016}
\bibinfo{author}{H.~Leppinen},
\newblock \bibinfo{title}{{Deploying a single-launch nanosatellite
  constellation to several orbital planes using drag maneuvers}},
\newblock \bibinfo{journal}{Acta Astronautica} \bibinfo{volume}{121}
  (\bibinfo{year}{2016}) \bibinfo{pages}{23--28}.
  \DOIprefix\doi{10.1016/j.actaastro.2015.12.036}.
\bibitem[{Leonard et~al.(1989)Leonard, Hollister, and Bergmann}]{Leonard1989}
\bibinfo{author}{C.~Leonard}, \bibinfo{author}{W.~Hollister},
  \bibinfo{author}{E.~Bergmann},
\newblock \bibinfo{title}{{Orbital Formationkeeping with Differential Drag}},
\newblock \bibinfo{journal}{Journal of Guidance, Control, and Dynamics}
  \bibinfo{volume}{12} (\bibinfo{year}{1989}) \bibinfo{pages}{108--113}.
  \DOIprefix\doi{10.2514/3.20374}.
\bibitem[{Traub et~al.(2019)Traub, Romano, Binder, Boxberger, Herdrich,
  Fasoulas, Roberts, Smith, Edmondson, Haigh, Crisp, Oiko, Lyons, Worrall,
  Livadiotti, Becedas, Gonz{\'{a}}lez, Dominguez, Gonz{\'{a}}lez, Ghizoni,
  Jungnell, Bay, Morsb{\o}l, Garcia-Almi{\~{n}}ana, Rodriguez-Donaire, Sureda,
  Kataria, Outlaw, Villain, Perez, Conte, Belkouchi, Schwalber, and
  Hei{\ss}erer}]{Traub2019a}
\bibinfo{author}{C.~Traub}, \bibinfo{author}{F.~Romano},
  \bibinfo{author}{T.~Binder}, \bibinfo{author}{A.~Boxberger},
  \bibinfo{author}{G.~H. Herdrich}, \bibinfo{author}{S.~Fasoulas},
  \bibinfo{author}{P.~C. Roberts}, \bibinfo{author}{K.~L. Smith},
  \bibinfo{author}{S.~Edmondson}, \bibinfo{author}{S.~J. Haigh},
  \bibinfo{author}{N.~H. Crisp}, \bibinfo{author}{V.~T.~A. Oiko},
  \bibinfo{author}{R.~E. Lyons}, \bibinfo{author}{S.~D. Worrall},
  \bibinfo{author}{S.~Livadiotti}, \bibinfo{author}{J.~Becedas},
  \bibinfo{author}{G.~Gonz{\'{a}}lez}, \bibinfo{author}{R.~M. Dominguez},
  \bibinfo{author}{D.~Gonz{\'{a}}lez}, \bibinfo{author}{L.~Ghizoni},
  \bibinfo{author}{V.~Jungnell}, \bibinfo{author}{K.~Bay},
  \bibinfo{author}{J.~Morsb{\o}l}, \bibinfo{author}{D.~Garcia-Almi{\~{n}}ana},
  \bibinfo{author}{S.~Rodriguez-Donaire}, \bibinfo{author}{M.~Sureda},
  \bibinfo{author}{D.~Kataria}, \bibinfo{author}{R.~Outlaw},
  \bibinfo{author}{R.~Villain}, \bibinfo{author}{J.~S. Perez},
  \bibinfo{author}{A.~Conte}, \bibinfo{author}{B.~Belkouchi},
  \bibinfo{author}{A.~Schwalber}, \bibinfo{author}{B.~Hei{\ss}erer},
\newblock \bibinfo{title}{{On the exploitation of differential aerodynamic lift
  and drag as a means to control satellite formation flight}},
\newblock \bibinfo{journal}{CEAS Space Journal}  (\bibinfo{year}{2019}).
  \URLprefix \url{http://link.springer.com/10.1007/s12567-019-00254-y}.
  \DOIprefix\doi{10.1007/s12567-019-00254-y}.
\bibitem[{Bevilacqua and Romano(2008)}]{Bevilacqua2008}
\bibinfo{author}{R.~Bevilacqua}, \bibinfo{author}{M.~Romano},
\newblock \bibinfo{title}{{Rendezvous Maneuvers of Multiple Spacecraft Using
  Differential Drag Under J2 Perturbation}},
\newblock \bibinfo{journal}{Journal of Guidance, Control, and Dynamics}
  \bibinfo{volume}{31} (\bibinfo{year}{2008}) \bibinfo{pages}{1595--1607}.
  \DOIprefix\doi{10.2514/1.36362}.
\bibitem[{Dell'Elce and Kerschen(2015)}]{DellElce2015}
\bibinfo{author}{L.~Dell'Elce}, \bibinfo{author}{G.~Kerschen},
\newblock \bibinfo{title}{{Optimal propellantless rendez-vous using
  differential drag}},
\newblock \bibinfo{journal}{Acta Astronautica} \bibinfo{volume}{109}
  (\bibinfo{year}{2015}) \bibinfo{pages}{112--123}. \URLprefix
  \url{http://dx.doi.org/10.1016/j.actaastro.2015.01.011}.
  \DOIprefix\doi{10.1016/j.actaastro.2015.01.011}.
\bibitem[{{Virgili Llop} et~al.(2015{\natexlab{a}}){Virgili Llop}, Roberts,
  Palmer, Hobbs, and Kingston}]{VirgiliLlop2015}
\bibinfo{author}{J.~{Virgili Llop}}, \bibinfo{author}{P.~C. Roberts},
  \bibinfo{author}{K.~Palmer}, \bibinfo{author}{S.~E. Hobbs},
  \bibinfo{author}{J.~Kingston},
\newblock \bibinfo{title}{{Descending Sun-Synchronous Orbits with Aerodynamic
  Inclination Correction}},
\newblock \bibinfo{journal}{Journal of Guidance, Control, and Dynamics}
  \bibinfo{volume}{38} (\bibinfo{year}{2015}{\natexlab{a}})
  \bibinfo{pages}{831--842}. \URLprefix
  \url{http://arc.aiaa.org/doi/10.2514/1.G000183}.
  \DOIprefix\doi{10.2514/1.G000183}.
\bibitem[{{Virgili Llop} et~al.(2015{\natexlab{b}}){Virgili Llop}, Roberts, and
  Hara}]{VirgiliLlop2015a}
\bibinfo{author}{J.~{Virgili Llop}}, \bibinfo{author}{P.~C. Roberts},
  \bibinfo{author}{N.~C. Hara},
\newblock \bibinfo{title}{{Atmospheric Interface Reentry Point Targeting Using
  Aerodynamic Drag Control}},
\newblock \bibinfo{journal}{Journal of Guidance, Control, and Dynamics}
  \bibinfo{volume}{38} (\bibinfo{year}{2015}{\natexlab{b}})
  \bibinfo{pages}{403--413}. \URLprefix
  \url{http://arc.aiaa.org/doi/abs/10.2514/1.G000884}.
  \DOIprefix\doi{10.2514/1.G000884}.
\bibitem[{Omar and Bevilacqua(2019)}]{Omar2019}
\bibinfo{author}{S.~Omar}, \bibinfo{author}{R.~Bevilacqua},
\newblock \bibinfo{title}{{Guidance, navigation, and control solutions for
  spacecraft re-entry point targeting using aerodynamic drag}},
\newblock \bibinfo{journal}{Acta Astronautica} \bibinfo{volume}{155}
  (\bibinfo{year}{2019}) \bibinfo{pages}{389--405}. \URLprefix
  \url{https://doi.org/10.1016/j.actaastro.2018.10.016
  https://linkinghub.elsevier.com/retrieve/pii/S0094576518302893}.
  \DOIprefix\doi{10.1016/j.actaastro.2018.10.016}.
\bibitem[{Gargasz(2007)}]{Gargasz2007}
\bibinfo{author}{M.~L. Gargasz}, \bibinfo{title}{{Optimal Spacecraft Attitude
  Control Using Aerodynamic Torques}}, \bibinfo{type}{Msc thesis}, Air Force
  Institute of Technology, \bibinfo{year}{2007}.
\bibitem[{Auret and Steyn(2011)}]{Auret2011}
\bibinfo{author}{J.~Auret}, \bibinfo{author}{W.~H. Steyn},
\newblock \bibinfo{title}{{Design of an Aerodynamic Attitude Control System for
  a Cubesat}},
\newblock in: \bibinfo{booktitle}{62nd International Astronautical Congress},
  \bibinfo{address}{Cape Town, South Africa}, \bibinfo{year}{2011}.
\bibitem[{Mostaza-Prieto and Roberts(2017)}]{Mostaza-Prieto2017}
\bibinfo{author}{D.~Mostaza-Prieto}, \bibinfo{author}{P.~C. Roberts},
\newblock \bibinfo{title}{{Perigee Attitude Maneuvers of Geostationary
  Satellites During Electric Orbit Raising}},
\newblock \bibinfo{journal}{Journal of Guidance, Control, and Dynamics}
  (\bibinfo{year}{2017}) \bibinfo{pages}{1--12}. \URLprefix
  \url{https://arc.aiaa.org/doi/10.2514/1.G002370}.
  \DOIprefix\doi{10.2514/1.G002370}.
\bibitem[{Livadiotti et~al.(2021)Livadiotti, Crisp, Roberts, Oiko, Christensen,
  Dominguez, and Herdrich}]{Livadiotti2021}
\bibinfo{author}{S.~Livadiotti}, \bibinfo{author}{N.~Crisp},
  \bibinfo{author}{P.~Roberts}, \bibinfo{author}{V.~Oiko},
  \bibinfo{author}{S.~Christensen}, \bibinfo{author}{R.~M. Dominguez},
  \bibinfo{author}{G.~Herdrich},
\newblock \bibinfo{title}{{Uncertainties and Design of Active Aerodynamic
  Attitude Control in Very Low Earth Orbit}},
\newblock \bibinfo{journal}{Journal of Guidance, Control, and Dynamics (under
  review)}  (\bibinfo{year}{2021}).
\bibitem[{Mishne and Edlerman(2017)}]{Mishne2017}
\bibinfo{author}{D.~Mishne}, \bibinfo{author}{E.~Edlerman},
\newblock \bibinfo{title}{{Collision-avoidance maneuver of satellites using
  drag and solar radiation pressure}},
\newblock \bibinfo{journal}{Journal of Guidance, Control, and Dynamics}
  \bibinfo{volume}{40} (\bibinfo{year}{2017}) \bibinfo{pages}{1191--1205}.
  \URLprefix \url{https://arc.aiaa.org/doi/10.2514/1.G002376}.
  \DOIprefix\doi{10.2514/1.G002376}.
\bibitem[{Huang et~al.(2017)Huang, Yan, and Zhou}]{Huang2017}
\bibinfo{author}{X.~Huang}, \bibinfo{author}{Y.~Yan},
  \bibinfo{author}{Y.~Zhou},
\newblock \bibinfo{title}{{Underactuated spacecraft formation reconfiguration
  with collision avoidance}},
\newblock \bibinfo{journal}{Acta Astronautica} \bibinfo{volume}{131}
  (\bibinfo{year}{2017}) \bibinfo{pages}{166--181}. \URLprefix
  \url{http://dx.doi.org/10.1016/j.actaastro.2016.11.037}.
  \DOIprefix\doi{10.1016/j.actaastro.2016.11.037}.
\bibitem[{Maclay and Tuttle(2005)}]{Maclay2005}
\bibinfo{author}{T.~D. Maclay}, \bibinfo{author}{C.~Tuttle},
\newblock \bibinfo{title}{{Satellite Stationkeeping of the ORBCOMM
  Constellation Via Active Control of Atmospheric Drag: Operations,
  Constraints, and Performance}},
\newblock \bibinfo{journal}{Advances in the Astronautical Sciences}
  \bibinfo{volume}{120} (\bibinfo{year}{2005}).
\bibitem[{Gangestad et~al.(2013)Gangestad, Hardy, and Hinkley}]{Gangestad2013}
\bibinfo{author}{J.~W. Gangestad}, \bibinfo{author}{B.~S. Hardy},
  \bibinfo{author}{D.~A. Hinkley},
\newblock \bibinfo{title}{{Operations, Orbit Determination, and Formation
  Control of the AeroCube-4 CubeSats}},
\newblock in: \bibinfo{booktitle}{27th Annual AIAA/USU Conference on Small
  Satellites}, \bibinfo{publisher}{American Institute of Aeronautics and
  Astronautics (AIAA)}, \bibinfo{address}{Logan, UT}, \bibinfo{year}{2013}.
\bibitem[{Foster et~al.(2018)Foster, Mason, Vittaldev, Leung, Beukelaers,
  Stepan, and Zimmerman}]{Foster2018}
\bibinfo{author}{C.~Foster}, \bibinfo{author}{J.~Mason},
  \bibinfo{author}{V.~Vittaldev}, \bibinfo{author}{L.~Leung},
  \bibinfo{author}{V.~Beukelaers}, \bibinfo{author}{L.~Stepan},
  \bibinfo{author}{R.~Zimmerman},
\newblock \bibinfo{title}{{Constellation phasing with differential drag on
  planet labs satellites}},
\newblock \bibinfo{journal}{Journal of Spacecraft and Rockets}
  \bibinfo{volume}{55} (\bibinfo{year}{2018}) \bibinfo{pages}{473--483}.
  \DOIprefix\doi{10.2514/1.A33927}.
\bibitem[{Tossman et~al.(1980)Tossman, Mobley, Fountain, Heffernan, Ray, and
  Williams}]{Tossman1980}
\bibinfo{author}{B.~Tossman}, \bibinfo{author}{F.~Mobley},
  \bibinfo{author}{G.~Fountain}, \bibinfo{author}{K.~Heffernan},
  \bibinfo{author}{J.~Ray}, \bibinfo{author}{C.~Williams},
\newblock \bibinfo{title}{{MAGSAT attitude control system design and
  performance}},
\newblock in: \bibinfo{booktitle}{Guidance and Control Conference},
  \bibinfo{publisher}{American Institute of Aeronautics and Astronautics
  (AIAA)}, \bibinfo{address}{Danvers, MA}, \bibinfo{year}{1980}, pp.
  \bibinfo{pages}{95--104}. \URLprefix
  \url{http://arc.aiaa.org/doi/10.2514/6.1980-1730}.
  \DOIprefix\doi{10.2514/6.1980-1730}.
\bibitem[{Stengle(1980)}]{Stengle1980}
\bibinfo{author}{T.~H. Stengle},
\newblock \bibinfo{title}{{MagSat Attitude Dynamics and Control: Some
  Observations and Explanations}},
\newblock in: \bibinfo{editor}{J.~Teles} (Ed.), \bibinfo{booktitle}{Fiifth
  Annual Flight Mechanics/Estimation Theory Symposium},
  \bibinfo{address}{Greenbelt, MD}, \bibinfo{year}{1980}, pp.
  \bibinfo{pages}{1--30}.
\bibitem[{Wertz and Larson(1999)}]{Wertz1999}
\bibinfo{editor}{J.~R. Wertz}, \bibinfo{editor}{W.~J. Larson} (Eds.),
  \bibinfo{title}{{Space Mission Analysis and Design}}, \bibinfo{edition}{3}
  ed., \bibinfo{publisher}{Microcosm Press/Kluwer Academic Publishers},
  \bibinfo{address}{El Segundo, CA}, \bibinfo{year}{1999}.
\bibitem[{Wertz et~al.(2011)Wertz, Everett, and Puschell}]{Wertz2011}
\bibinfo{editor}{J.~R. Wertz}, \bibinfo{editor}{D.~F. Everett},
  \bibinfo{editor}{J.~J. Puschell} (Eds.), \bibinfo{title}{{Space Mission
  Engineering: The New SMAD}}, \bibinfo{edition}{1} ed.,
  \bibinfo{publisher}{Microcosm Press}, \bibinfo{address}{Hawthorne, CA},
  \bibinfo{year}{2011}.
\bibitem[{Fortescue et~al.(2011)Fortescue, Swinerd, and Stark}]{Fortescue2011}
\bibinfo{editor}{P.~Fortescue}, \bibinfo{editor}{G.~Swinerd},
  \bibinfo{editor}{J.~Stark} (Eds.), \bibinfo{title}{{Spacecraft Systems
  Engineering}}, \bibinfo{edition}{4} ed., \bibinfo{publisher}{John Wiley {\&}
  Sons, Ltd.}, \bibinfo{address}{Chichester, UK}, \bibinfo{year}{2011}.
\bibitem[{Lambe and Martins(2012)}]{Lambe2012}
\bibinfo{author}{A.~B. Lambe}, \bibinfo{author}{J.~R. R.~A. Martins},
\newblock \bibinfo{title}{{Extensions to the design structure matrix for the
  description of multidisciplinary design, analysis, and optimization
  processes}},
\newblock \bibinfo{journal}{Structural and Multidisciplinary Optimization}
  \bibinfo{volume}{46} (\bibinfo{year}{2012}) \bibinfo{pages}{273--284}.
  \URLprefix \url{http://link.springer.com/10.1007/s00158-012-0763-y}.
  \DOIprefix\doi{10.1007/s00158-012-0763-y}.
\bibitem[{Mosher(1999)}]{Mosher1999}
\bibinfo{author}{T.~J. Mosher},
\newblock \bibinfo{title}{{Conceptual Spacecraft Design Using a Genetic
  Algorithm Trade Selection Process}},
\newblock \bibinfo{journal}{Journal of Aircraft} \bibinfo{volume}{36}
  (\bibinfo{year}{1999}) \bibinfo{pages}{200--208}.
  \DOIprefix\doi{10.2514/2.2426}.
\bibitem[{Fukunaga et~al.(1997)Fukunaga, Chien, Mutz, Sherwood, and
  Stechert}]{Fukunaga1997}
\bibinfo{author}{A.~S. Fukunaga}, \bibinfo{author}{S.~Chien},
  \bibinfo{author}{D.~Mutz}, \bibinfo{author}{R.~L. Sherwood},
  \bibinfo{author}{A.~D. Stechert},
\newblock \bibinfo{title}{{Automating the process of optimization in spacecraft
  design}},
\newblock \bibinfo{journal}{IEEE Aerospace Applications Conference Proceedings}
  \bibinfo{volume}{4} (\bibinfo{year}{1997}) \bibinfo{pages}{411--427}.
  \DOIprefix\doi{10.1109/aero.1997.577524}.
\bibitem[{Barnhart et~al.(2009)Barnhart, Kichkaylo, and Hoag}]{Barnhart2009a}
\bibinfo{author}{D.~J. Barnhart}, \bibinfo{author}{T.~Kichkaylo},
  \bibinfo{author}{L.~Hoag},
\newblock \bibinfo{title}{{SPIDR: Integrated Systems Engineering
  Design-to-Simulation Software for Satellite Build}},
\newblock in: \bibinfo{booktitle}{Proceedings of the Conference on Systems
  Engineering Research (CSER)}, \bibinfo{address}{Loughborough, UK},
  \bibinfo{year}{2009}.
\bibitem[{Lowe and Macdonald(2014)}]{Lowe2014}
\bibinfo{author}{C.~Lowe}, \bibinfo{author}{M.~Macdonald},
\newblock \bibinfo{title}{{Rapid model-based inter-disciplinary design of a
  CubeSat mission}},
\newblock \bibinfo{journal}{Acta Astronautica} \bibinfo{volume}{105}
  (\bibinfo{year}{2014}) \bibinfo{pages}{321--332}. \URLprefix
  \url{https://linkinghub.elsevier.com/retrieve/pii/S0094576514003695}.
  \DOIprefix\doi{10.1016/j.actaastro.2014.10.002}.
\bibitem[{Hwang et~al.(2014)Hwang, Lee, Cutler, and Martins}]{Hwang2014}
\bibinfo{author}{J.~T. Hwang}, \bibinfo{author}{D.~Y. Lee},
  \bibinfo{author}{J.~W. Cutler}, \bibinfo{author}{J.~R. Martins},
\newblock \bibinfo{title}{{Large-Scale Multidisciplinary Optimization of a
  Small Satellite's Design and Operation}},
\newblock \bibinfo{journal}{Journal of Spacecraft and Rockets}
  \bibinfo{volume}{51} (\bibinfo{year}{2014}) \bibinfo{pages}{1648--1663}.
  \DOIprefix\doi{10.2514/1.A32751}.
\bibitem[{{Le Moigne} et~al.(2017){Le Moigne}, Dabney, de~Weck, Foreman,
  Grogan, Holland, Hughes, and Nag}]{LeMoigne2017}
\bibinfo{author}{J.~{Le Moigne}}, \bibinfo{author}{P.~Dabney},
  \bibinfo{author}{O.~L. de~Weck}, \bibinfo{author}{V.~Foreman},
  \bibinfo{author}{P.~Grogan}, \bibinfo{author}{M.~Holland},
  \bibinfo{author}{S.~P. Hughes}, \bibinfo{author}{S.~Nag},
\newblock \bibinfo{title}{{Tradespace analysis tool for designing
  constellations (TAT-C)}},
\newblock in: \bibinfo{booktitle}{IEEE International Geoscience and Remote
  Sensing Symposium (IGARSS)}, \bibinfo{publisher}{IEEE},
  \bibinfo{address}{Fort Worth, TX}, \bibinfo{year}{2017}. \URLprefix
  \url{http://ieeexplore.ieee.org/document/8127168/}.
  \DOIprefix\doi{10.1109/IGARSS.2017.8127168}.
\bibitem[{Ferringer and Spencer(2006)}]{Ferringer2006}
\bibinfo{author}{M.~P. Ferringer}, \bibinfo{author}{D.~B. Spencer},
\newblock \bibinfo{title}{{Satellite Constellation Design Tradeoffs Using
  Multiple-Objective Evolutionary Computation}},
\newblock \bibinfo{journal}{Journal of Spacecraft and Rockets}
  \bibinfo{volume}{43} (\bibinfo{year}{2006}) \bibinfo{pages}{1404--1411}.
  \DOIprefix\doi{10.2514/1.18788}.
\bibitem[{de~Weck et~al.(2004)de~Weck, de~Neufville, and Chaize}]{DeWeck2004}
\bibinfo{author}{O.~L. de~Weck}, \bibinfo{author}{R.~de~Neufville},
  \bibinfo{author}{M.~Chaize},
\newblock \bibinfo{title}{{Staged Deployment of Communications Satellite
  Constellations in Low Earth Orbit}},
\newblock \bibinfo{journal}{Journal of Aerospace Computing, Information, and
  Communication} \bibinfo{volume}{1} (\bibinfo{year}{2004})
  \bibinfo{pages}{119--136}. \DOIprefix\doi{10.2514/1.6346}.
\bibitem[{Crisp et~al.(2019)Crisp, Smith, and Hollingsworth}]{Crisp2019}
\bibinfo{author}{N.~H. Crisp}, \bibinfo{author}{K.~L. Smith},
  \bibinfo{author}{P.~M. Hollingsworth},
\newblock \bibinfo{title}{{An integrated design methodology for the deployment
  of constellations of small satellites}},
\newblock \bibinfo{journal}{The Aeronautical Journal} \bibinfo{volume}{123}
  (\bibinfo{year}{2019}) \bibinfo{pages}{1193--1215}. \URLprefix
  \url{https://www.cambridge.org/core/product/identifier/S0001924019000575/type/journal{\_}article}.
  \DOIprefix\doi{10.1017/aer.2019.57}.
\bibitem[{Spangelo and Longmier(2014)}]{Spangelo2014}
\bibinfo{author}{S.~C. Spangelo}, \bibinfo{author}{B.~W. Longmier},
\newblock \bibinfo{title}{{Small Spacecraft System-level Design and
  Optimization for Interplanetary Trajectories}},
\newblock in: \bibinfo{booktitle}{AIAA/AAS Astrodynamics Specialist
  Conference}, \bibinfo{number}{August}, \bibinfo{publisher}{American Institute
  of Aeronautics and Astronautics}, \bibinfo{address}{San Diego, CA},
  \bibinfo{year}{2014}, pp. \bibinfo{pages}{1--15}. \URLprefix
  \url{http://arc.aiaa.org/doi/10.2514/6.2014-4125}.
  \DOIprefix\doi{10.2514/6.2014-4125}.
\bibitem[{Dono et~al.(2018)Dono, Plice, Mueting, Conn, and Ho}]{Dono2018}
\bibinfo{author}{A.~Dono}, \bibinfo{author}{L.~Plice},
  \bibinfo{author}{J.~Mueting}, \bibinfo{author}{T.~Conn},
  \bibinfo{author}{M.~Ho},
\newblock \bibinfo{title}{{Propulsion trade studies for spacecraft swarm
  mission design}},
\newblock in: \bibinfo{booktitle}{2018 IEEE Aerospace Conference},
  \bibinfo{publisher}{IEEE}, \bibinfo{address}{Big Sky, MT},
  \bibinfo{year}{2018}, pp. \bibinfo{pages}{1--12}. \URLprefix
  \url{https://ieeexplore.ieee.org/document/8396492/}.
  \DOIprefix\doi{10.1109/AERO.2018.8396492}.
\bibitem[{Krejci and Lozano(2018)}]{Krejci2018}
\bibinfo{author}{D.~Krejci}, \bibinfo{author}{P.~C. Lozano},
\newblock \bibinfo{title}{{Space Propulsion Technology for Small Spacecraft}},
\newblock \bibinfo{journal}{Proceedings of the IEEE} \bibinfo{volume}{106}
  (\bibinfo{year}{2018}) \bibinfo{pages}{362--378}. \URLprefix
  \url{http://ieeexplore.ieee.org/document/8252908/}.
  \DOIprefix\doi{10.1109/JPROC.2017.2778747}.
\bibitem[{de~Weck and Chang(2002)}]{DeWeck2002}
\bibinfo{author}{O.~L. de~Weck}, \bibinfo{author}{D.~Chang},
\newblock \bibinfo{title}{{Architecture Trade Methodology for LEO Personal
  Communication Systems}},
\newblock in: \bibinfo{booktitle}{20th International Communicatuions Satellite
  Systems Conference}, \bibinfo{publisher}{American Institute of Aeronautics
  and Astronautics (AIAA)}, \bibinfo{address}{Montreal, Canada},
  \bibinfo{year}{2002}. \DOIprefix\doi{10.2514/6.2002-1866}.
\bibitem[{Fearn(2005)}]{Fearn2005}
\bibinfo{author}{D.~G. Fearn},
\newblock \bibinfo{title}{{Economical remote sensing from a low altitude with
  continuous drag compensation}},
\newblock \bibinfo{journal}{Acta Astronautica} \bibinfo{volume}{56}
  (\bibinfo{year}{2005}) \bibinfo{pages}{555--572}. \URLprefix
  \url{https://linkinghub.elsevier.com/retrieve/pii/S0094576504003704}.
  \DOIprefix\doi{10.1016/j.actaastro.2004.09.052}.
\bibitem[{Shao et~al.(2016)Shao, Madni, and Wertz}]{Shao2016}
\bibinfo{author}{A.~E. Shao}, \bibinfo{author}{A.~M. Madni},
  \bibinfo{author}{J.~R. Wertz},
\newblock \bibinfo{title}{{Quantifying the Effect of Orbit Altitude on Mission
  Cost for Earth Observation Satellites}},
\newblock in: \bibinfo{booktitle}{54th AIAA Aerospace Sciences Meeting},
  \bibinfo{number}{January}, \bibinfo{publisher}{American Institute of
  Aeronautics and Astronautics}, \bibinfo{address}{Reston, Virginia},
  \bibinfo{year}{2016}, pp. \bibinfo{pages}{1--13}. \URLprefix
  \url{http://arc.aiaa.org/doi/10.2514/6.2016-0974}.
  \DOIprefix\doi{10.2514/6.2016-0974}.
\bibitem[{{Virgili Llop} et~al.(2014){Virgili Llop}, Roberts, Hao, {Ramio
  Tomas}, and Beauplet}]{VirgiliLlop2014a}
\bibinfo{author}{J.~{Virgili Llop}}, \bibinfo{author}{P.~C. Roberts},
  \bibinfo{author}{Z.~Hao}, \bibinfo{author}{L.~{Ramio Tomas}},
  \bibinfo{author}{V.~Beauplet},
\newblock \bibinfo{title}{{Very Low Earth Orbit mission concepts for Earth
  Observation: Benefits and challenges}},
\newblock in: \bibinfo{booktitle}{12th Reinventing Space Conference},
  \bibinfo{address}{London, UK}, \bibinfo{year}{2014}.
\bibitem[{Bacon and Olivier(2017)}]{Bacon2017}
\bibinfo{author}{A.~Bacon}, \bibinfo{author}{B.~Olivier},
\newblock \bibinfo{title}{{Skimsats: bringing down the cost of Earth
  Observation}},
\newblock in: \bibinfo{editor}{S.~Hatton} (Ed.),
  \bibinfo{booktitle}{Proceedings of the 12th Reinventing Space Conference},
  \bibinfo{publisher}{Springer International Publishing},
  \bibinfo{address}{Cham}, \bibinfo{year}{2017}, pp. \bibinfo{pages}{1--7}.
  \URLprefix \url{http://link.springer.com/10.1007/978-3-319-34024-1{\_}1}.
  \DOIprefix\doi{10.1007/978-3-319-34024-1_1}.
\bibitem[{McCreary(2019)}]{McCreary2019}
\bibinfo{author}{L.~McCreary},
\newblock \bibinfo{title}{{A satellite mission concept for high drag
  environments}},
\newblock \bibinfo{journal}{Aerospace Science and Technology}
  \bibinfo{volume}{92} (\bibinfo{year}{2019}) \bibinfo{pages}{972--989}.
  \URLprefix \url{https://doi.org/10.1016/j.ast.2019.06.033}.
  \DOIprefix\doi{10.1016/j.ast.2019.06.033}.
\bibitem[{Bertolucci et~al.(2020)Bertolucci, Barato, Toson, and
  Pavarin}]{Bertolucci2020}
\bibinfo{author}{G.~Bertolucci}, \bibinfo{author}{F.~Barato},
  \bibinfo{author}{E.~Toson}, \bibinfo{author}{D.~Pavarin},
\newblock \bibinfo{title}{{Impact of propulsion system characteristics on the
  potential for cost reduction of earth observation missions at very low
  altitudes}},
\newblock \bibinfo{journal}{Acta Astronautica} \bibinfo{volume}{176}
  (\bibinfo{year}{2020}) \bibinfo{pages}{173--191}. \URLprefix
  \url{https://doi.org/10.1016/j.actaastro.2020.06.018
  https://linkinghub.elsevier.com/retrieve/pii/S0094576520303854}.
  \DOIprefix\doi{10.1016/j.actaastro.2020.06.018}.
\bibitem[{Nishiyama(2003)}]{Nishiyama2003}
\bibinfo{author}{K.~Nishiyama},
\newblock \bibinfo{title}{{Air Breathing Ion Engine Concept}},
\newblock in: \bibinfo{booktitle}{54th International Astronautical Congress},
  \bibinfo{publisher}{American Institute of Aeronautics and Astronautics
  (AIAA)}, \bibinfo{address}{Bremen, Germany}, \bibinfo{year}{2003}. \URLprefix
  \url{http://arc.aiaa.org/doi/10.2514/6.IAC-03-S.4.02}.
  \DOIprefix\doi{10.2514/6.IAC-03-S.4.02}.
\bibitem[{Hisamoto et~al.(2012)Hisamoto, Nishiyama, and
  Kunninaka}]{Hisamoto2012}
\bibinfo{author}{Y.~Hisamoto}, \bibinfo{author}{K.~Nishiyama},
  \bibinfo{author}{H.~Kunninaka},
\newblock \bibinfo{title}{{Design of Air Intake for Air Breathing Ion Engine}},
\newblock in: \bibinfo{booktitle}{63rd International Astronautical Congress},
  \bibinfo{publisher}{International Astronautical Federation (IAF)},
  \bibinfo{address}{Naples, Italy}, \bibinfo{year}{2012}.
\bibitem[{{Di Cara} et~al.(2007){Di Cara}, {Gonzalez del Amo}, Santovicenzo,
  {Carnicero Dominguez}, Arcioni, Caldwell, and Roma}]{DiCara2007}
\bibinfo{author}{D.~{Di Cara}}, \bibinfo{author}{J.~{Gonzalez del Amo}},
  \bibinfo{author}{A.~Santovicenzo}, \bibinfo{author}{B.~{Carnicero
  Dominguez}}, \bibinfo{author}{M.~Arcioni}, \bibinfo{author}{A.~Caldwell},
  \bibinfo{author}{I.~Roma},
\newblock \bibinfo{title}{{RAM Electric Propulsion for Low Earth Orbit
  Operation: an ESA study}},
\newblock in: \bibinfo{booktitle}{30th IEPC - International Electric Propulsion
  Conference}, \bibinfo{address}{Florence, Italy}, \bibinfo{year}{2007}, pp.
  \bibinfo{pages}{1--8}.
\bibitem[{Romano et~al.(2018)Romano, Massuti-Ballester, Binder, Herdrich,
  Fasoulas, and Sch{\"{o}}nherr}]{Romano2018c}
\bibinfo{author}{F.~Romano}, \bibinfo{author}{B.~Massuti-Ballester},
  \bibinfo{author}{T.~Binder}, \bibinfo{author}{G.~H. Herdrich},
  \bibinfo{author}{S.~Fasoulas}, \bibinfo{author}{T.~Sch{\"{o}}nherr},
\newblock \bibinfo{title}{{System analysis and test-bed for an
  atmosphere-breathing electric propulsion system using an inductive plasma
  thruster}},
\newblock \bibinfo{journal}{Acta Astronautica} \bibinfo{volume}{147}
  (\bibinfo{year}{2018}) \bibinfo{pages}{114--126}. \URLprefix
  \url{https://doi.org/10.1016/j.actaastro.2018.03.031
  https://linkinghub.elsevier.com/retrieve/pii/S009457651730406X}.
  \DOIprefix\doi{10.1016/j.actaastro.2018.03.031}.
\bibitem[{Romano et~al.(2015)Romano, Binder, Herdrich, Fasoulas, and
  Sch{\"{o}}nherr}]{Romano2015}
\bibinfo{author}{F.~Romano}, \bibinfo{author}{T.~Binder},
  \bibinfo{author}{G.~H. Herdrich}, \bibinfo{author}{S.~Fasoulas},
  \bibinfo{author}{T.~Sch{\"{o}}nherr},
\newblock \bibinfo{title}{{Air-Intake Design Investigation for an
  Air-Breathing}},
\newblock in: \bibinfo{booktitle}{34th IEPC - International Electric Propulsion
  Conference}, \bibinfo{address}{Kobe, Japan}, \bibinfo{year}{2015}, pp.
  \bibinfo{pages}{1--27}.
\bibitem[{Romano et~al.(2016)Romano, Binder, Herdrich, Fasoulas, and
  Sch{\"{o}}nherr}]{Romano2016}
\bibinfo{author}{F.~Romano}, \bibinfo{author}{T.~Binder},
  \bibinfo{author}{G.~H. Herdrich}, \bibinfo{author}{S.~Fasoulas},
  \bibinfo{author}{T.~Sch{\"{o}}nherr},
\newblock \bibinfo{title}{{Intake Design for an Atmosphere-Breathing Electric
  Propulsion System}},
\newblock in: \bibinfo{booktitle}{Space Propulsion}, \bibinfo{number}{May},
  \bibinfo{address}{Rome, Italy}, \bibinfo{year}{2016}.
\bibitem[{Binder et~al.(2016)Binder, Boldini, Romano, Herdrich, and
  Fasoulas}]{Binder2016}
\bibinfo{author}{T.~Binder}, \bibinfo{author}{P.~Boldini},
  \bibinfo{author}{F.~Romano}, \bibinfo{author}{G.~H. Herdrich},
  \bibinfo{author}{S.~Fasoulas},
\newblock \bibinfo{title}{{Transmission probabilities of rarefied flows in the
  application of atmosphere-breathing electric propulsion}},
\newblock in: \bibinfo{booktitle}{AIP Conference Proceedings}, volume
  \bibinfo{volume}{1786}, \bibinfo{year}{2016}. \URLprefix
  \url{http://aip.scitation.org/doi/abs/10.1063/1.4967689}.
  \DOIprefix\doi{10.1063/1.4967689}.
\bibitem[{Crisp et~al.(2018)Crisp, Livadiotti, and Roberts}]{Crisp2018}
\bibinfo{author}{N.~H. Crisp}, \bibinfo{author}{S.~Livadiotti},
  \bibinfo{author}{P.~C. Roberts},
\newblock \bibinfo{title}{{A Semi-Analytical Method for Calculating Revisit
  Time for Satellite Constellations with Discontinuous Coverage}},
\newblock \bibinfo{journal}{arXiv e-prints}  (\bibinfo{year}{2018}). \URLprefix
  \url{http://arxiv.org/abs/1807.02021}.
  \href{http://arxiv.org/abs/1807.02021}{{\tt arXiv:1807.02021}}.
\bibitem[{Vallado(2013)}]{Vallado2013}
\bibinfo{author}{D.~A. Vallado}, \bibinfo{title}{{Fundamentals of Astrodynamics
  and Applications}}, \bibinfo{edition}{4} ed., \bibinfo{publisher}{Microcosm
  Press/Springer}, \bibinfo{address}{Hawthorne, CA}, \bibinfo{year}{2013}.
\bibitem[{Votel and Sinclair(2012)}]{Votel2012}
\bibinfo{author}{R.~Votel}, \bibinfo{author}{D.~Sinclair},
\newblock \bibinfo{title}{{Comparison of Control Moment Gyros and Reaction
  Wheels for Small Earth-Observing Satellites}},
\newblock in: \bibinfo{booktitle}{26th Annual AIAA/USU Conference on Small
  Satellites}, \bibinfo{publisher}{American Institute of Aeronautics and
  Astronautics (AIAA)}, \bibinfo{address}{Logan, UT}, \bibinfo{year}{2012}.
\bibitem[{Humble et~al.(1995)Humble, Henry, and Larson}]{Humble1995}
\bibinfo{editor}{R.~W. Humble}, \bibinfo{editor}{G.~N. Henry},
  \bibinfo{editor}{W.~J. Larson} (Eds.), \bibinfo{title}{{Space Propulsion
  Analysis and Design}}, \bibinfo{edition}{1} ed.,
  \bibinfo{publisher}{McGraw-Hill}, \bibinfo{address}{New York, NY},
  \bibinfo{year}{1995}.
\bibitem[{Chiasson and Lozano(2012)}]{Chiasson2012}
\bibinfo{author}{T.~M. Chiasson}, \bibinfo{author}{P.~C. Lozano},
  \bibinfo{title}{{Modeling the Characteristics of Propulsion Systems Providing
  Less Than 10 N Thrust}}, \bibinfo{type}{Msc.}, Massachusetts Institute of
  Technology, \bibinfo{year}{2012}.
\bibitem[{Mahr et~al.(2020)Mahr, Tu, and Gupta}]{Mahr2020}
\bibinfo{author}{E.~Mahr}, \bibinfo{author}{A.~Tu}, \bibinfo{author}{A.~Gupta},
\newblock \bibinfo{title}{{Development of the Small Satellite Cost Model 2019
  (SSCM19)}},
\newblock in: \bibinfo{booktitle}{2020 IEEE Aerospace Conference},
  \bibinfo{publisher}{IEEE}, \bibinfo{address}{Big Sky, MT},
  \bibinfo{year}{2020}, pp. \bibinfo{pages}{1--12}. \URLprefix
  \url{https://ieeexplore.ieee.org/document/9172374/}.
  \DOIprefix\doi{10.1109/AERO47225.2020.9172374}.
\bibitem[{Walsh et~al.(2020)Walsh, Berthoud, and Allen}]{Walsh2020}
\bibinfo{author}{J.~Walsh}, \bibinfo{author}{L.~Berthoud},
  \bibinfo{author}{C.~Allen},
\newblock \bibinfo{title}{{Drag reduction through shape optimisation for
  satellites in Very Low Earth Orbit}},
\newblock \bibinfo{journal}{Acta Astronautica}  (\bibinfo{year}{2020}).
  \URLprefix
  \url{https://linkinghub.elsevier.com/retrieve/pii/S0094576520305579}.
  \DOIprefix\doi{10.1016/j.actaastro.2020.09.018}.
\bibitem[{Costes et~al.(2012)Costes, Cassar, and Escarrat}]{Costes2012}
\bibinfo{author}{V.~Costes}, \bibinfo{author}{G.~Cassar},
  \bibinfo{author}{L.~Escarrat},
\newblock \bibinfo{title}{{Optical design of a compact telescope for the next
  generation Earth observation system}},
\newblock in: \bibinfo{booktitle}{International Conference on Space Optics —
  ICSO}, volume \bibinfo{volume}{10564}, \bibinfo{publisher}{SPIE},
  \bibinfo{year}{2012}. \URLprefix
  \url{https://www.spiedigitallibrary.org/conference-proceedings-of-spie/10564/2309055/Optical-design-of-a-compact-telescope-for-the-next-generation/10.1117/12.2309055.full}.
  \DOIprefix\doi{10.1117/12.2309055}.
\bibitem[{Metwally et~al.(2020)Metwally, Bazan, and Eltohamy}]{Metwally2020}
\bibinfo{author}{M.~Metwally}, \bibinfo{author}{T.~M. Bazan},
  \bibinfo{author}{F.~Eltohamy},
\newblock \bibinfo{title}{{Design of Very High-Resolution Satellite Telescopes
  Part I: Optical System Design}},
\newblock \bibinfo{journal}{IEEE Transactions on Aerospace and Electronic
  Systems} \bibinfo{volume}{56} (\bibinfo{year}{2020})
  \bibinfo{pages}{1202--1208}. \URLprefix
  \url{https://ieeexplore.ieee.org/document/8768011/}.
  \DOIprefix\doi{10.1109/TAES.2019.2929969}.
\bibitem[{Tomiyasu(1978)}]{Tomiyasu1978}
\bibinfo{author}{K.~Tomiyasu},
\newblock \bibinfo{title}{{Tutorial Review of Synthetic-Aperture Radar (SAR)
  with Applications to Imaging of the Ocean Surface}},
\newblock \bibinfo{journal}{Proceedings of the IEEE} \bibinfo{volume}{66}
  (\bibinfo{year}{1978}) \bibinfo{pages}{563--583}.
  \DOIprefix\doi{10.1109/PROC.1978.10961}.
\bibitem[{Cutrona(1990)}]{Cutrona1990}
\bibinfo{author}{L.~Cutrona},
\newblock \bibinfo{title}{{Synthetic Aperture Radar}},
\newblock in: \bibinfo{editor}{M.~I. Skolnik} (Ed.), \bibinfo{booktitle}{Radar
  Handbook}, \bibinfo{edition}{2} ed., \bibinfo{publisher}{McGraw-Hill},
  \bibinfo{address}{New York, NY}, \bibinfo{year}{1990}.
\bibitem[{Freeman(2018)}]{Freeman2018}
\bibinfo{author}{A.~Freeman},
\newblock \bibinfo{title}{{Design Principles for Smallsat SARs}},
\newblock in: \bibinfo{booktitle}{32nd Annual AIAA/USU Conference on Small
  Satellites}, \bibinfo{address}{Logan, UT}, \bibinfo{year}{2018}.
\bibitem[{Maxwell(1878)}]{March1878}
\bibinfo{author}{J.~C. Maxwell},
\newblock \bibinfo{title}{{III. On stresses in rarefied gases arising from
  inequalities of temperature}},
\newblock \bibinfo{journal}{Proceedings of the Royal Society of London}
  \bibinfo{volume}{27} (\bibinfo{year}{1878}) \bibinfo{pages}{304--308}.
  \URLprefix
  \url{https://royalsocietypublishing.org/doi/10.1098/rspl.1878.0052}.
  \DOIprefix\doi{10.1098/rspl.1878.0052}.
\bibitem[{{AIAA Atmospheric and Space Environments Comimittee on
  Standards}(2010)}]{AIAA2010}
\bibinfo{author}{{AIAA Atmospheric and Space Environments Comimittee on
  Standards}}, \bibinfo{title}{{Guide to Reference and Standard Atmosphere
  Models}}, \bibinfo{type}{AIAA G-003C-2010(2016)}, American Institude of
  Aeronautics and Astronautics, \bibinfo{address}{Reston, VA},
  \bibinfo{year}{2010}.
\bibitem[{{ISO/TC 20/SC 14 Space systems and operations}(2013)}]{ISO2013}
\bibinfo{author}{{ISO/TC 20/SC 14 Space systems and operations}},
  \bibinfo{title}{{ISO 14222:2013 Space environment (natural and artificial)
  — Earth upper atmosphere}}, \bibinfo{type}{Standard}, International
  Organisation for Standardization, \bibinfo{address}{Geneva, Switzerland},
  \bibinfo{year}{2013}.
\bibitem[{Mostaza-Prieto(2017)}]{Mostaza-Prieto2017a}
\bibinfo{author}{D.~Mostaza-Prieto}, \bibinfo{title}{{Characterisation and
  Applications of Aerodynamic Torques on Satellites}}, \bibinfo{type}{Phd
  thesis}, The University of Manchester, \bibinfo{year}{2017}.
\bibitem[{Sinpetru et~al.(2021)Sinpetru, Crisp, Mostaza-Prieto, Livadiotti, and
  Roberts}]{Sinpetru2021a}
\bibinfo{author}{L.~A. Sinpetru}, \bibinfo{author}{N.~H. Crisp},
  \bibinfo{author}{D.~Mostaza-Prieto}, \bibinfo{author}{S.~Livadiotti},
  \bibinfo{author}{P.~C.~E. Roberts},
\newblock \bibinfo{title}{{ADBSat: Methodology of a novel panel method tool for
  aerodynamic analysis of satellites}},
\newblock \bibinfo{journal}{Submitted to Computer Physics Communications}
  (\bibinfo{year}{2021}). \URLprefix \url{http://arxiv.org/abs/2104.05543}.
  \href{http://arxiv.org/abs/2104.05543}{{\tt arXiv:2104.05543}}.
\bibitem[{Goebel and Katz(2008)}]{Goebel2008}
\bibinfo{author}{D.~M. Goebel}, \bibinfo{author}{I.~Katz},
  \bibinfo{title}{{Fundamentals of Electric Propulsion: Ion and Hall
  Thrusters}}, \bibinfo{publisher}{John Wiley {\&} Sons, Inc.},
  \bibinfo{address}{Hoboken, NJ, USA}, \bibinfo{year}{2008}. \URLprefix
  \url{http://doi.wiley.com/10.1002/9780470436448}.
  \DOIprefix\doi{10.1002/9780470436448}.
\bibitem[{Holste et~al.(2020)Holste, Dietz, Scharmann, Keil, Henning,
  Zsch{\"{a}}tzsch, Reitemeyer, Nausch{\"{u}}tt, Kiefer, Kunze, Zorn, Heiliger,
  Joshi, Probst, Th{\"{u}}ringer, Volkmar, Packan, Peterschmitt, Brinkmann,
  Zaunick, Thoma, Kretschmer, Leiter, Schippers, Hannemann, and
  Klar}]{Holste2020}
\bibinfo{author}{K.~Holste}, \bibinfo{author}{P.~Dietz},
  \bibinfo{author}{S.~Scharmann}, \bibinfo{author}{K.~Keil},
  \bibinfo{author}{T.~Henning}, \bibinfo{author}{D.~Zsch{\"{a}}tzsch},
  \bibinfo{author}{M.~Reitemeyer}, \bibinfo{author}{B.~Nausch{\"{u}}tt},
  \bibinfo{author}{F.~Kiefer}, \bibinfo{author}{F.~Kunze},
  \bibinfo{author}{J.~Zorn}, \bibinfo{author}{C.~Heiliger},
  \bibinfo{author}{N.~Joshi}, \bibinfo{author}{U.~Probst},
  \bibinfo{author}{R.~Th{\"{u}}ringer}, \bibinfo{author}{C.~Volkmar},
  \bibinfo{author}{D.~Packan}, \bibinfo{author}{S.~Peterschmitt},
  \bibinfo{author}{K.~T. Brinkmann}, \bibinfo{author}{H.-G. Zaunick},
  \bibinfo{author}{M.~H. Thoma}, \bibinfo{author}{M.~Kretschmer},
  \bibinfo{author}{H.~J. Leiter}, \bibinfo{author}{S.~Schippers},
  \bibinfo{author}{K.~Hannemann}, \bibinfo{author}{P.~J. Klar},
\newblock \bibinfo{title}{{Ion thrusters for electric propulsion: Scientific
  issues developing a niche technology into a game changer}},
\newblock \bibinfo{journal}{Review of Scientific Instruments}
  \bibinfo{volume}{91} (\bibinfo{year}{2020}) \bibinfo{pages}{061101}.
  \URLprefix \url{http://aip.scitation.org/doi/10.1063/5.0010134}.
  \DOIprefix\doi{10.1063/5.0010134}.
\bibitem[{Hohman(2012)}]{Hohman2012}
\bibinfo{author}{K.~Hohman},
\newblock \bibinfo{title}{{Atmospheric breathing electric thruster for
  planetary exploration}},
\newblock in: \bibinfo{booktitle}{NIAC Spring Symposium},
  \bibinfo{address}{Pasadena, CA}, \bibinfo{year}{2012}.
\bibitem[{Takahashi(2019)}]{Takahashi2019}
\bibinfo{author}{K.~Takahashi},
\newblock \bibinfo{title}{{Helicon-type radiofrequency plasma thrusters and
  magnetic plasma nozzles}},
\newblock \bibinfo{journal}{Reviews of Modern Plasma Physics}
  \bibinfo{volume}{3} (\bibinfo{year}{2019}) \bibinfo{pages}{3}. \URLprefix
  \url{https://doi.org/10.1007/s41614-019-0024-2
  http://link.springer.com/10.1007/s41614-019-0024-2}.
  \DOIprefix\doi{10.1007/s41614-019-0024-2}.
\bibitem[{Romano et~al.(2021)Romano, Espinosa-Orozco, Herdrich, Crisp, Roberts,
  Holmes, Edmondson, Haigh, Livadiotti, Macario-Rojas, Oiko, Sinpetru, Smith,
  Becedas, Sulliotti-Linner, Bisgaard, Christensen, Hanessian, Jensen, Nielsen,
  Chan, Fasoulas, Traub, Garcia-Almi{\~{n}}ana, Rodr{\'{i}}guez-Donaire,
  Sureda, Kataria, Belkouchi, Conte, Seminari, and Villain}]{Romano2021}
\bibinfo{author}{F.~Romano}, \bibinfo{author}{J.~Espinosa-Orozco},
  \bibinfo{author}{G.~Herdrich}, \bibinfo{author}{N.~H. Crisp},
  \bibinfo{author}{P.~C. Roberts}, \bibinfo{author}{B.~E. Holmes},
  \bibinfo{author}{S.~Edmondson}, \bibinfo{author}{S.~Haigh},
  \bibinfo{author}{S.~Livadiotti}, \bibinfo{author}{A.~Macario-Rojas},
  \bibinfo{author}{V.~Oiko}, \bibinfo{author}{L.~Sinpetru},
  \bibinfo{author}{K.~Smith}, \bibinfo{author}{J.~Becedas},
  \bibinfo{author}{V.~Sulliotti-Linner}, \bibinfo{author}{M.~Bisgaard},
  \bibinfo{author}{S.~Christensen}, \bibinfo{author}{V.~Hanessian},
  \bibinfo{author}{T.~Jensen}, \bibinfo{author}{J.~Nielsen},
  \bibinfo{author}{Y.-A. Chan}, \bibinfo{author}{S.~Fasoulas},
  \bibinfo{author}{C.~Traub}, \bibinfo{author}{D.~Garcia-Almi{\~{n}}ana},
  \bibinfo{author}{S.~Rodr{\'{i}}guez-Donaire}, \bibinfo{author}{M.~Sureda},
  \bibinfo{author}{D.~Kataria}, \bibinfo{author}{B.~Belkouchi},
  \bibinfo{author}{A.~Conte}, \bibinfo{author}{S.~Seminari},
  \bibinfo{author}{R.~Villain},
\newblock \bibinfo{title}{{Intake Design for an Atmosphere-Breathing Electric
  Propulsion System (ABEP)}},
\newblock \bibinfo{journal}{Acta Astronautica (under review)}
  (\bibinfo{year}{2021}).
\bibitem[{Foreman et~al.(2016)Foreman, {Le Moigne}, and {De
  Weck}}]{Foreman2016}
\bibinfo{author}{V.~L. Foreman}, \bibinfo{author}{J.~{Le Moigne}},
  \bibinfo{author}{O.~{De Weck}},
\newblock \bibinfo{title}{{A Survey of Cost Estimating Methodologies for
  Distributed Spacecraft Missions}},
\newblock in: \bibinfo{booktitle}{AIAA SPACE 2016},
  \bibinfo{publisher}{American Institute of Aeronautics and Astronautics
  (AIAA)}, \bibinfo{address}{Long Beach, CA}, \bibinfo{year}{2016}, pp.
  \bibinfo{pages}{1--15}. \URLprefix
  \url{http://arc.aiaa.org/doi/10.2514/6.2016-5245}.
  \DOIprefix\doi{10.2514/6.2016-5245}.
\bibitem[{Paek et~al.(2020)Paek, Balasubramanian, Kim, and de~Weck}]{Paek2020}
\bibinfo{author}{S.~W. Paek}, \bibinfo{author}{S.~Balasubramanian},
  \bibinfo{author}{S.~Kim}, \bibinfo{author}{O.~de~Weck},
\newblock \bibinfo{title}{{Small-Satellite Synthetic Aperture Radar for
  Continuous Global Biospheric Monitoring: A Review}},
\newblock \bibinfo{journal}{Remote Sensing} \bibinfo{volume}{12}
  (\bibinfo{year}{2020}) \bibinfo{pages}{2546}. \URLprefix
  \url{https://www.mdpi.com/2072-4292/12/16/2546}.
  \DOIprefix\doi{10.3390/rs12162546}.

\end{thebibliography}

\end{document}